\documentclass[10pt]{article}
\usepackage[utf8]{inputenc}
\usepackage[a4paper,top=3cm,bottom=4cm,left=3cm,right=3cm]{geometry}
\usepackage[linktocpage=true,colorlinks,citecolor=blue,linkcolor=blue,urlcolor=blue]{hyperref}
\usepackage{cite}
\usepackage{graphicx}
\usepackage{xcolor}
\usepackage{amsmath,amssymb,amsfonts}
\usepackage{mathtools,mathrsfs}
\usepackage{float}
\usepackage{comment}
\usepackage[default]{gillius}
\usepackage{setspace}
\usepackage{enumitem}
\usepackage{array}
\usepackage{breqn}

\newcommand{\T}{\mathbb{T}}
\newcommand{\Q}{\mathbb{Q}}
\newcommand{\E}{\mathcal{E}}

\newcommand{\C}{\mathcal{C}}

\newcommand{\R}{\mathcal{R}}
\newcommand{\D}{\mathcal{D}}

\newcommand{\Lie}{\mathcal{L}}
\newcommand{\K}{\mathcal{K}}
\newcommand{\dd}{\text{d}}

\newcommand{\Chi}{\text{X}}
\newcommand{\HH}{\mathcal{H}}

\newcommand{\LCBig}[2]{\ensuremath{\genfrac{\{}{\}}{0pt}{1}{#1}{#2}}}
\def\g{\mathfrak{g}}
\def\d{\mathfrak{d}}

\DeclareRobustCommand{\rchi}{{\mathpalette\irchi\relax}}
\newcommand{\irchi}[2]{\raisebox{\depth}{$#1\chi$}}

\begin{document}

\title{Cosmological teleparallel perturbations}

\begin{center}
\Large{Cosmological teleparallel perturbations}\\
\end{center}
\begin{center}
    \renewcommand*{\thefootnote}{\fnsymbol{footnote}}
    Lavinia Heisenberg\footnote{Electronic address: \href{mailto:lavinia.heisenberg@phys.ethz.ch}{L.Heisenberg@ThPhys.Uni-Heidelberg.DE}}\\
    \textit{Institut für Theoretische Physik, Philosophenweg 16, 69120 Heidelberg, Germany}\\
\vspace*{.5cm}
    Manuel Hohmann\footnote{Electronic address: \href{mailto:manuel.hohmann@ut.ee}{manuel.hohmann@ut.ee}}\\
    \textit{Laboratory of Theoretical Physics, Institute of Physics, University of Tartu, W. Ostwaldi 1, 50411 Tartu, Estonia}\\
\vspace*{.5cm}
    Simon Kuhn\footnote{Electronic address: \href{mailto:simkuhn@phys.ethz.ch}{simkuhn@phys.ethz.ch}}\\
    \textit{Institute for Theoretical Physics, ETH Zurich, Wolfgang-Pauli-Strasse 27, 8093 Zurich, Switzerland}
\end{center}
\setcounter{footnote}{0}

\begin{abstract}
There has been growing interest in $f(\Q)$ gravity, which has led to significant advancements in the field. However, it is important to note that most studies in this area were based on the coincident gauge, thus overlooking the impact of the connection degrees of freedom. In this work, we pay special attention to the connection when studying perturbations in general teleparallel, metric teleparallel, and symmetric teleparallel theories of gravity. We do not just examine perturbations in the metric, but also in the affine connection. To illustrate this, we investigate cosmological perturbations in $f(G)$, $f(\T)$, and $f(\Q)$ gravity with and without matter in form of an additional scalar field for spatially flat and curved FLRW geometries. Our perturbative analysis reveals that for general $f(\Q)$ backgrounds, there are up to seven degrees of freedom, depending on the background connection. This is in perfect agreement with the upper bound on degrees of freedom established for the first time in \href{https://doi.org/10.1002/prop.202300185}{Fortschr. Phys. 2023, 2300185}.
 In $f(G)$ and $f(\T)$ gravity theories, only two tensor modes propagate in the gravity sector on generic curved cosmological backgrounds, indicating strong coupling problems.
In the context of $f(\Q)$ cosmology, we find that for a particular background connection, where all seven modes propagate, there is at least one ghost degree of freedom. 
For all other choices of the connection the ghost can be avoided at the cost of strong coupling problem, where only four degrees of freedom propagate. Hence, all of the cosmologies within the teleparallel families of theories in form of $f(G)$, $f(\T)$, and $f(\Q)$ suffer either from strong coupling or from ghost instabilities. A direct coupling of the matter field to the connection or non-minimal couplings might alter these results. 
\end{abstract}

\newpage
\tableofcontents

\newpage
\section{Introduction}


Over the past century, General Relativity (GR) has garnered an unequivocal empirical foundation across an extensive range of scales, verified through high precision laboratory experiments and triumphantly elucidating gravitational phenomena at macroscopic scales, exemplified by the $\Lambda$CDM model's remarkable predictions concerning the universe's evolution and the emergence of cosmic structures \cite{Will:2005va}. Pure GR masterfully anticipates not only highly symmetric backgrounds but also perturbations atop these backgrounds, encompassing diverse phenomena like temperature fluctuations in the CMB, matter perturbations engendering galaxy formation, and gravitational waves \cite{Mukhanov:2005sc, LIGOScientific:2016aoc}. Nonetheless, from its inception, the theory has been marred by the enigmatic absence of gravitating vacuum energy, leading to the cosmological constant problem \cite{RevModPhys.61.1}. This intriguing quandary serves as a compelling indication that Einstein's theory might not comprehensively account for gravity in the infrared regime.

In the realm of contemporary scientific inquiries, GR faces intriguing challenges that warrant comprehensive exploration and resolution. These challenges encompass the enigmatic phenomena of singularities, the complexities surrounding cosmological tensions, and the pressing quest to reconcile gravity with the realm of quantum mechanics. Moreover, the compelling evidence supporting the universe's current accelerating expansion \cite{SupernovaCosmologyProject:1998vns,SupernovaSearchTeam:1998fmf} intensifies the quest for a viable generalization of the theory of gravity on cosmological scales. The overarching goal is to explore the prospect of incorporating dynamical dark energy, which could potentially serve as a mechanism to address the cosmological constant issue in tandem with explaining the accelerating cosmic expansion.

A plethora of consistent extensions for GR have been proposed, each with its unique approach \cite{Clifton:2011jh, Copeland:2006wr, Heisenberg:2018vsk}. Given that GR stands as the singular effective field theory of a massless spin 2 degree of freedom in four dimensions, any extension of the theory inevitably introduces supplementary degrees of freedom. Within the framework of field theory, these novel degrees of freedom often take the form of additional scalar, vector, or tensor fields \cite{Heisenberg:2018vsk, Heisenberg:2014rta, Heisenberg:2018acv, Deffayet:2009wt, Horndeski:1974ux}. Hence, their absence in the perturbation analysis would be a direct indication to strong coupling problems. In the realm of geometrical frameworks, recent interest has been directed towards teleparallel theories. These theories diverge from traditional approaches by not solely relying on the spacetime metric but also incorporating a curvature-free connection to formulate the action. The versatility of teleparallel theories arises from their ability to decompose arbitrary connections into components such as the Levi-Cevita connection, fundamental in constructing the Einstein-Hilbert action, alongside a distinct torsion and non-metricity contribution. Consequently, teleparallel theories do not find their basis in curvature but rather pivot on torsion and non-metricity exclusively \cite{Heisenberg:2018vsk, BeltranJimenez:2019, Hohmann:2022mlc}. Within this context, one has the option to retain both torsion and non-metricity \cite{BeltranJimenez:2019odq}, or alternatively, to impose additional restrictions, leading to metric-teleparallel theories, which rely solely on torsion, or symmetric-teleparallel theories, which exclusively depend on non-metricity \cite{BeltranJimenez:2017}. In each of these three scenarios, it is possible to formulate theories that dynamically mirror GR, denoted as teleparallel (TEGR) and symmetric teleparallel equivalents to GR (STEGR) \cite{BeltranJimenez:2018vdo, BeltranJimenez:2019}.

The various representations of GR, encompassing curvature, torsion, and non-metricity, provide distinct and viable foundations for theories of modified gravity \cite{Heisenberg:2018vsk, Heisenberg:2023lru}. Notably, these foundations offer the flexibility of elevating the corresponding Lagrangians to arbitrary non-linear functions. Recent scholarly literature has witnessed significant attention directed towards $f(\Q)$ theories of modified gravity \cite{BeltranJimenez:2017, Heisenberg:2018vsk, Heisenberg:2023lru}, where scalar functions of the non-metricity tensor are utilized. These $f(\Q)$ theories have proven to be particularly intriguing, exhibiting noteworthy cosmological \cite{BeltranJimenez:2019tme, DAmbrosio:2020nev, Bajardi:2020wl, Ayuso:2020dcu, Frusciante:2021sio, Anagnostopoulos:2021ydo, Atayde:2021pgb, DAmbrosio:2021pnd, Hohmann:2021, Capozziello:2022wgl, Dimakis:2022rkd, Esposito:2022omp} and astrophysical \cite{Zhao:2021zab, Lin:2021uqa, DAmbrosio:2021zpm, Banerjee:2021mqk, Wang:2021zaz, Parsaei:2022wnu, Maurya:2022wwa} implications.

In this paper, we endeavor to expand the scope of cosmological teleparallel models by incorporating first-order perturbations. Our investigation encompasses perturbations not only in the metric but also in the teleparallel connection. As the connection is required to be flat, and in the cases of metric or symmetric teleparallism, metric-compatible or torsion-free, respectively, the perturbations must adhere to specific forms, which we elucidate and explore, including their corresponding gauge transformations. These perturbations are subsequently categorized by the scalar-vector-tensor decomposition, which divides them into modes that behave either as scalars, vectors, or tensors under spatial coordinate transformation. These different types of perturbations effectively decouple at first order perturbation theory. For further reference and comparison, we also invite the reader to consult \cite{Hohmann:2020vcv}. We rigorously calculate the actions of perturbations within the contexts of general teleparallel, metric teleparallel, and symmetric teleparallel theories. As illustrative examples, we apply these actions to $f(G)$, $f(\T)$, and $f(\Q)$ theories, which, notably, do not share equivalence with $f(R)$ gravity. To cover a wide range of cosmological backgrounds we consider both spatially flat and curved geometries, as well as pure vacuum with the cosmological expansion driven by gravity alone, as well as with a matter source in the form of a scalar field with arbitrary potential. A particular focus is placed on discerning the number of propagating gravitational perturbation variables, an aspect which, in GR, is solely represented by the two tensor modes.

The paper's structure unfolds as follows: In Section 2, we provide an introduction to teleparallel gravity and establish the conventions used throughout the analysis. Moving on to Section 3, we comprehensively outline the complete perturbations of both metric and flat connections, along with their corresponding gauge transformations and scalar-vector-tensor decompositions. Sections 4 to 6 delve into in-depth discussions of perturbations in $f(G)$, $f(\T)$, and $f(\Q)$ cosmologies, respectively. Lastly, Section 7 is dedicated to discussing and interpreting the results obtained from the study.

Throughout the paper we adopt natural units where $8\pi G=c=1$. Initially, we maintain the spacetime dimension $d$ as a general variable. However, for the subsequent calculations in sections 4 to 6, we specifically focus on $d=4$ dimensions.

\section{Teleparallel gravity theories}
Teleparallel gravity theories are alternative approaches to describe gravity. These theories utilize the notion of a curvature-free connection, focusing on torsion and non-metricity as fundamental aspects. Unlike in GR, where gravity is attributed to the curvature of spacetime, in teleparallel theories, it arises from the torsion or non-metricity of spacetime. We consider metric-affine theories on a $d$ dimensional manifold $\mathcal{M}$, i.e. beside the metric tensor $g_{\mu\nu}$ we also have an affine connection $\Gamma^\alpha{}_{\mu\nu}$ associated to a covariant derivative $\nabla_\mu$. The affine connection is in general not the Levi-Cevita connection $\LCBig{\alpha}{\mu\nu}$ and is not associated to the torsion-free and metric-compatible covariant derivative $\mathcal{D}_\mu$. One can show \cite{Jimenez:2018} that the general affine connection may be decomposed as
\begin{equation}\label{eq:ConnectionDecomp}
    \Gamma^\alpha{}_{\mu\nu}=\LCBig{\alpha}{\mu\nu}+M^\alpha{}_{\mu\nu}\,,\,M^\alpha{}_{\mu\nu}=K^\alpha{}_{\mu\nu}+L^\alpha{}_{\mu\nu}\,,
\end{equation}
with the torsion and non-metricity tensors\footnote{We define symmetrization and antisymmetrization by
\begin{align}
    2A_{(\mu\nu)}&=A_{\mu\nu}+A_{\nu\mu}\,,\nonumber\\
    2A_{[\mu\nu]}&=A_{\mu\nu}-A_{\nu\mu}\,,\nonumber
\end{align}
for any tensor $A_{\mu\nu}$.}
\begin{align}
    T^\alpha{}_{\mu\nu}&=2\Gamma^\alpha{}_{[\mu\nu]}\,,\\
    Q_{\alpha\mu\nu}&=\nabla_\alpha g_{\mu\nu}=\partial_\alpha g_{\mu\nu}-2\Gamma^\lambda{}_{\alpha(\mu}g_{\nu)\lambda}\ ,
\end{align}
defining the contortion and disformation tensors
\begin{align}
    2K^\alpha{}_{\mu\nu}&=T^\alpha{}_{\mu\nu}+2T_{(\mu}{}^\alpha{}_{\nu)}\,,\\
    2L^\alpha{}_{\mu\nu}&=Q^\alpha{}_{\mu\nu}-2Q_{(\mu}{}^\alpha{}_{\nu)}\,.
\end{align}
We define the Riemann tensors associated to $\Gamma^\alpha{}_{\mu\nu}$ and the Levi-Cevita connection by
\begin{align}
    R^\alpha{}_{\beta\mu\nu}&=2\partial_{[\mu}\Gamma^\alpha{}_{\nu]\beta}+2\Gamma^\alpha{}_{[\mu|\lambda|}\Gamma^\lambda{}_{\nu]\beta}\ ,\\
    \R^\alpha{}_{\beta\mu\nu}&=2\partial_{[\mu}\LCBig{\alpha}{\nu]\beta}+2\LCBig{\alpha}{[\mu|\lambda|}\LCBig{\lambda}{\nu]\beta}\,,
\end{align}
respectively. We then have the identities
\begin{align}
    &[\nabla_\mu,\nabla_\nu]\omega=-T^\lambda{}_{\mu\nu}\nabla_\lambda\omega\,,\\
    &[\nabla_\mu,\nabla_\nu]W^\alpha=R^\alpha{}_{\beta\mu\nu}W^\beta-T^\lambda{}_{\mu\nu}\nabla_\lambda W^\alpha\,,
\end{align}
with $\omega$ a scalar field and $W^\alpha$ a vector field. From this, one can derive the Bianchi identities
\begin{align}
    R^\mu{}_{[\alpha\beta\gamma]}-\nabla_{[\alpha}T^\mu{}_{\beta\gamma]}+T^\nu{}_{[\alpha\beta}T^\mu{}_{\gamma]\nu}&=0\,,\label{eq:BianchiI}\\
    \nabla_{[\alpha}R^\mu{}_{|\nu|\beta\gamma]}-T^\lambda{}_{[\alpha\beta}R^\mu{}_{|\nu|\gamma]\lambda}&=0\,.
\end{align}
As usual we define the Ricci tensor and scalar by $R_{\mu\nu}=R^\lambda{}_{\mu\lambda\nu}$ and $R=R^\mu{}_\mu$, and similarly for the Riemann tensor of the Levi-Cevita connection. By plugging \eqref{eq:ConnectionDecomp} in the Riemann tensor one can show \cite{BeltranJimenez:2019odq, Heisenberg:2022mbo}
\begin{equation}
    R=\R+2\D_{[\mu}M^\mu{}_{\nu]}{}^\nu+2M^\mu{}_{[\mu|\lambda|}M^\lambda{}_{\nu]}{}^\nu\,.
\end{equation}
We can now impose the teleparallel condition
\begin{equation}
    R^\alpha{}_{\beta\mu\nu}=0
\end{equation}
on the affine connection such that
\begin{equation}\label{eq:TPFundEqu}
    \R=2\D_{[\nu}M^\mu{}_{\mu]}{}^\nu+2M^\mu{}_{[\nu|\lambda|}M^\lambda{}_{\mu]}{}^\nu\,.
\end{equation}
This allows us to define a metric-affine theory with the action \cite{BeltranJimenez:2019odq}
\begin{equation}\label{eq:ActionGTEGR}
    S=\int\dd^dx(\sqrt{-g}\,G+\lambda_\alpha{}^{\beta\mu\nu}R^\alpha{}_{\beta\mu\nu})\,,
\end{equation}
with
\begin{align}
    G&=2M^\mu{}_{[\nu|\lambda|}M^\lambda{}_{\mu]}{}^\nu=-\frac14T^\mu{}_{\nu\lambda}T_\mu{}^{\nu\lambda}-\frac12 T^\mu{}_{\nu\lambda}T^\nu{}_\mu{}^\lambda+T^\mu T_\mu-\nonumber\\
    &-\frac14 Q_{\mu\nu\lambda}Q^{\mu\nu\lambda}+\frac12Q_{\mu\nu\lambda}Q^{\nu\mu\lambda}+\frac14 Q_\mu Q^\mu-\frac12Q_\mu\tilde Q^\mu-\\
    &-T^\mu{}_{\nu\lambda}Q^\nu{}_\mu{}^\lambda+T^\mu Q_\mu-T^\mu\tilde Q_\mu\,,\nonumber
\end{align}
where we defined the traces
\begin{equation}
    T_\mu=T^\nu{}_{\mu\nu}\,,\,Q^\mu=Q^\mu{}_\nu{}^\nu\,,\,\tilde Q^\mu=Q^{\nu\mu}{}_\nu\,.
\end{equation}
$\lambda_\alpha{}^{\beta\mu\nu}$ is a Lagrange multiplier ensuring that the connection is flat.  According to \eqref{eq:TPFundEqu}, this theory is dynamically equivalent to GR, wherein the affine connection trivializes at the level of the equations of motion, as GR is unaffected by it.

Additionally, assuming the connection to be metric-compatible as well, i.e.
\begin{align}
 \,Q_{\alpha\mu\nu}=0\,,
\end{align}
one can define the torsion scalar
\begin{equation}\label{eq:MTFundEqu}
    \T=\frac14T^\mu{}_{\nu\lambda}T_\mu{}^{\nu\lambda}-\frac12 T^\mu{}_{\nu\lambda}T^\nu{}_\mu{}^\lambda+T^\mu T_\mu\,,
\end{equation}
such that the theory given by the action
\begin{equation}\label{eq:ActionMTEGR}
    S=\int\dd^dx(\sqrt{-g}\,\T+\lambda_\alpha{}^{\beta\mu\nu}R^\alpha{}_{\beta\mu\nu}+\lambda^{\alpha\mu\nu}Q_{\alpha\mu\nu})
\end{equation}
is the teleparallel equivalent to GR. $\lambda^{\alpha\mu\nu}$ is another Lagrange multiplier that makes the connection metric-compatible. The theory is built purely out of the metric and the torsion tensor. Such theories are called metric-teleparallel or TEGR.

Imposing the affine connection to be teleparallel and torsion-free,
\begin{align}
    R^\alpha{}_{\beta\mu\nu}=0\,,\,T^\alpha{}_{\mu\nu}=0\,,
\end{align}
and defining the non-metricity scalar
\begin{equation}\label{eq:STFundEqu}
    \Q=-\frac14 Q_{\mu\nu\lambda}Q^{\mu\nu\lambda}+\frac12Q_{\mu\nu\lambda}Q^{\nu\mu\lambda}+\frac14 Q_\mu Q^\mu-\frac12Q_\mu\tilde Q^\mu\,,
\end{equation}
results in the action
\begin{equation}\label{eq:ActionSTEGR}
    S=\int\dd^dx(\sqrt{-g}\,\Q+\lambda_\alpha{}^{\beta\mu\nu}R^\alpha{}_{\beta\mu\nu}+\lambda_\alpha{}^{\mu\nu}T^\alpha{}_{\mu\nu})
\end{equation}
that is the symmetric teleparallel equivalent to GR \cite{BeltranJimenez:2017}. $\lambda_\alpha{}^{\mu\nu}$ is again another Lagrange multiplier ensuring now that the connection is torsion-free. These theories are constructed out of the metric and the non-metricity tensor and are called symmetric teleparallel or STEGR.\\

The fundamental underpinnings of these geometrical interpretations present a promising avenue for exploring modified gravity. GR's equivalent descriptions, encompassing curvature, torsion, and non-metricity, serve as distinct alternative frameworks, offering diverse starting points for theories of modified gravity. This is accomplished by elevating the corresponding scalar quantities, such as curvature, torsion, and non-metricity scalars, to arbitrary functions in the formulation of these modified gravity theories.
Simple examples are Lagrangians $\sqrt{-g}\Lie$ of the form $\Lie=f(G),f(\T),f(\Q)$ instead of $\Lie=G,\T,\Q$, where $f$ is a non-linear function \cite{Heisenberg:2018vsk}. Examples of how such theories affect the dynamics of spacetime in cosmological and black hole examples can be found in \cite{Hohmann:2017jao,DAmbrosio:2021,DAmbrosio2021:2109.04209v2,Heisenberg:2022mbo}.

In this paper we want to study the evolution of perturbations on top of a cosmological background spacetime in $f(G),f(\T),f(\Q)$ gravity theories, with a particular focus on $f(\Q)$. The cosmological expansion may then either be driven by the modified gravity, or the background matter. In order to obtain non-trivial background spacetimes we need additional matter, as we will see below, in most of our examples. We choose for simplicity a scalar field $\Xi$ for the matter field, with the matter action
\begin{equation}
    S_M=\int\dd^dx\,\sqrt{-g}\left(-\frac12 \nabla_\mu\Xi\nabla^\mu\Xi-\mathcal{V}(\Xi)\right)\,,
\end{equation}
where $\mathcal{V}$ is an unspecified potential. We only require the potential to vanish for $\Xi=0$, i.e. $\mathcal{V}(0)=0$, otherwise it will contribute in this case as an irrelevant cosmological constant. Consistency demands then also $\mathcal{V}'(0)=0$ to hold by the Klein-Gordon equation below. Since $\nabla_\mu\Xi=\mathcal{D}_\mu\Xi=\partial_\mu\Xi$, $S_M$ is independent of any connection, and thus the hypermomentum
\begin{equation}
\HH_\alpha{}^{\mu\nu}=-\frac{1}{\sqrt{-g}}\frac{\delta S_M}{\delta\Gamma^\alpha{}_{\mu\nu}}
\end{equation}
vanishes. The equation of motion of $\Xi$ is the Klein-Gordon equation in a curved spacetime,
\begin{equation}
    -\Box\Xi+\mathcal{V}'=0\,,
\end{equation}
with $\Box\equiv\mathcal{D}^\mu\mathcal{D}_\mu$. Primes on $\mathcal{V}$ are from now on derivatives w.r.t. its argument $\Xi$, i.e. $\mathcal{V}'=d\mathcal{V}/d\Xi$. The energy-momentum tensor of $\Xi$ is
\begin{equation}
    T_{\mu\nu}=-\frac{1}{\sqrt{-g}}\frac{\delta S_M}{\delta g^{\mu\nu}}=\frac12g_{\mu\nu}\left(-\frac12 \nabla_\rho\Xi\nabla^\rho\Xi-\mathcal{V}\right)+\frac12\nabla_\mu\Xi\nabla_\nu\Xi\,.
\end{equation}
The flexibility of choosing a potential $\mathcal{V}$ allows enough freedom to account for a wide range of cosmological background spacetimes including inflationary epochs with $\mathcal{V}$ approximately constant.

\section{Teleparallel perturbation theory}
Cosmological perturbation theory is a crucial tool in the study of the universe's large-scale structure and evolution. It investigates small deviations or fluctuations from the uniform background of the cosmos. By quantifying these perturbations in various components, such as matter and radiation, it enables the understanding of the formation of galaxies, cosmic structures, and the evolution of the universe. Cosmological perturbation theory plays a pivotal role in testing and constraining theoretical models, providing valuable insights into the fundamental nature of the chosen gravitational theory.

In this study, we delve into first-order cosmological perturbation theory within the realms of general teleparallel, metric-teleparallel, and symmetric teleparallel theories. Our particular focus lies on the non-equivalent cases compared to GR, specially within $f(\Q)$ gravity theories. As customary, we examine the metric perturbations
\begin{equation}
    g_{\mu\nu}\to g_{\mu\nu}+\delta g_{\mu\nu}\,,
\end{equation}
where $\delta g_{\mu\nu}$ encodes the $d(d+1)/2$ metric perturbations. Similarly, the affine connection is perturbed by
\begin{equation}
    \Gamma^\alpha{}_{\mu\nu}\to\Gamma^\alpha{}_{\mu\nu}+\delta\Gamma^\alpha{}_{\mu\nu}\,,
\end{equation}
with the $d^3$ connection perturbations $\delta\Gamma^\alpha{}_{\mu\nu}$. From now on $g_{\mu\nu},\Gamma^\alpha{}_{\mu\nu}$ refer to the background variables unless stated otherwise. Note that $\delta\Gamma^\alpha{}_{\mu\nu}$ is as a difference of connections a tensor.\\
\\
In teleparallel theories these perturbations propagate through the fundamental tensors like the Riemann, torsion, and non-metricity tensors as follows
\begin{align}
    \delta R^\alpha{}_{\beta\mu\nu}&=2\nabla_{[\mu}\delta\Gamma^\alpha{}_{\nu]\beta}+T^\lambda{}_{\mu\nu}\delta\Gamma^\alpha{}_{\lambda\beta}+2\delta\Gamma^\alpha{}_{[\mu|\lambda|}\delta\Gamma^\lambda{}_{\nu]\beta}\,,\\
    \delta T^\alpha{}_{\mu\nu}&=2\delta\Gamma^\alpha{}_{[\mu\nu]}\,,\\
    \delta Q_{\alpha\mu\nu}&=\nabla_\alpha\delta g_{\mu\nu}-2\delta\Gamma^\lambda{}_{\alpha(\mu}g_{\nu)\lambda}-2\delta\Gamma^\lambda{}_{\alpha(\mu}\delta g_{\nu)\lambda}\,.
\end{align}
Note that the perturbations of the Riemann and non-metricity tensors are quadratic in the perturbations, while the torsion tensor is linear. If we want the perturbed connection to still be teleparallel to first order we require
\begin{equation}
    \delta R^\alpha{}_{\beta\mu\nu}=2\nabla_{[\mu}\delta\Gamma^\alpha{}_{\nu]\beta}+T^\lambda{}_{\mu\nu}\delta\Gamma^\alpha{}_{\lambda\beta}=0\,,
\end{equation}
which can be solved by
\begin{equation}
    \delta\Gamma^\alpha{}_{\mu\nu}=\nabla_\mu\delta\Lambda^\alpha{}_\nu\,.
\end{equation}
The perturbations of the teleparallel connection are now encoded in the $d^2$ sized matrix $\delta\Lambda^\mu{}_\nu$.

In the case of metric-teleparallel theories we also need the non-metricity tensor to vanish to first order,
\begin{equation}
    \delta Q_{\alpha\mu\nu}=\nabla_\alpha \delta g_{\mu\nu}-2\nabla_\alpha\delta\Lambda^\lambda{}_{(\mu}\delta g_{\nu)\lambda}=\nabla_\alpha(\delta g_{\mu\nu}-2\delta\Lambda_{(\mu\nu)})=0\,,
\end{equation}
where we used that the background affine connection is metric-compatible. We thus have
\begin{equation}
    \delta\Lambda_{(\mu\nu)}=\frac12\delta g_{\mu\nu}\,,
\end{equation}
so that only $\delta\Lambda_{[\mu\nu]}$ are the $d(d-1)/2$ independent affine connection perturbations.

In the symmetric-teleparallel theories we need the connection perturbation to be symmetric in the two lower indices, $\nabla_{[\mu}\delta\Lambda^\alpha{}_{\nu]}=0$, from which it follows that
\begin{equation}
    \delta\Lambda^\mu{}_\nu=\nabla_\nu\delta\rchi^\mu\ ,
\end{equation}
where the $d$ perturbations of the affine connection are now given by $\delta\rchi^\mu$. Note that this gives the connection perturbations as a Lie derivative,
\begin{equation}
    \delta\Gamma^\alpha{}_{\mu\nu}=\nabla_\mu\nabla_\nu\delta\rchi^\alpha=\Lie_{\delta\rchi}\Gamma^\alpha{}_{\mu\nu}\,,
\end{equation}
where $\Lie_W$ is the Lie derivative w.r.t. a vector field $W^\mu$.

Alternatively one can also use the formulation of the flat connection in terms of tetrads, i.e. write the connection as
\begin{equation}
    \Gamma^\alpha{}_{\mu\nu}=e^\alpha{}_a\partial_\mu e_\nu{}^a
\end{equation}
with the tetrad $e_\mu{}^a$. Connection perturbations can then be given in terms of perturbations of the tetrad, $\delta e_\mu{}^a$, but we will stick to the perturbations $\delta\Gamma^\alpha{}_{\mu\nu}$ in this work. See appendix \ref{app:A} for more details on tetrad perturbations and their relations to our expressions.

Finally, for our choice of matter we have the perturbation of the scalar field
\begin{equation}
    \Xi\to\Xi+\delta\Xi\,,
\end{equation}
where $\delta\Xi$ is the single perturbation variable of the matter field. From now on we use $\Xi$ for the background scalar field.

\subsection{Gauge transformations}
Gauge transformations are fundamental in theoretical physics and represent changes in the mathematical description of a physical system that do not alter its observable quantities. In gauge theories, such as electromagnetism and GR, gauge transformations play a central role in maintaining the consistency and symmetry of the underlying physical laws.

Let us now consider an infinitesimal coordinate transformation $x^\mu\to x^\mu-\xi^\mu$, under which the metric variables change by
\begin{equation}
    \delta g_{\mu\nu}\to\delta g_{\mu\nu}+\Lie_\xi g_{\mu\nu}=\delta g_{\mu\nu}+2\D_{(\mu}\xi_{\nu)}\,.
\end{equation}
Similarly, the connection varies as \cite{BeltranJimenez:2020sih}
\begin{align}
    \delta\Gamma^\alpha{}_{\mu\nu}&\to\delta\Gamma^\alpha{}_{\mu\nu}+\Lie_\xi\Gamma^\alpha{}_{\mu\nu}=\nabla_\mu\delta\Lambda^\alpha{}_\nu+\nabla_\mu\nabla_\nu\xi^\alpha-\nabla_\mu(T^\alpha{}_{\nu\lambda}\xi^\lambda)-R^\alpha{}_{\nu\mu\lambda}\xi^\lambda=\\
    &=\nabla_\mu(\delta\Lambda^\alpha{}_\nu+\nabla_\nu\xi^\alpha-T^\alpha{}_{\nu\lambda}\xi^\lambda)\,,
\end{align}
so that we have the following gauge transformation of the teleparallel flat connection
\begin{equation}
    \delta\Lambda^\mu{}_\nu\to\delta\Lambda^\mu{}_\nu+\nabla_\nu\xi^\mu-T^\mu{}_{\nu\lambda}\xi^\lambda\,,
\end{equation}
which can be shown to be equivalent to
\begin{equation}
    \delta\Lambda_{\mu\nu}\to\delta\Lambda_{\mu\nu}+\mathcal{D}_\nu\xi_\mu+M_\mu{}^\lambda{}_\nu\xi_\lambda\,.
\end{equation}
From $K_{(\mu|\lambda|\nu)}=0$ it follows that for the symmetric part we have
\begin{equation}
    \delta\Lambda_{(\mu\nu)}\to\delta\Lambda_{(\mu\nu)}+\mathcal{D}_{(\mu}\xi_{\nu)}+L_{(\mu}{}^\lambda{}_{\nu)}\xi_\lambda\,,
\end{equation}
so in metric-teleparallel theories, where disformation vanishes, one has
\begin{equation}
    \delta\Lambda_{(\mu\nu)}\to\delta\Lambda_{(\mu\nu)}+\D_{(\mu}\xi_{\nu)}\,,
\end{equation}
which is compatible with $2\delta\Lambda_{(\mu\nu)}=\delta g_{\mu\nu}$. The antisymmetric part follows the gauge transformation
\begin{align}
    \delta\Lambda_{[\mu\nu]}&\to\delta\Lambda_{[\mu\nu]}+\partial_{[\nu}\xi_{\mu]}+K_\mu{}^\lambda{}_\nu\xi_\lambda+L_{[\mu}{}^\lambda{}_{\nu]}\xi_\lambda\,.
\end{align}
The gauge transformation of the connection perturbation thus also depends on the background metric and background connection; the symmetric part on the symmetric part of the disformation, and the antisymmetric part on the contortion and the antisymmetric part of the disformation.

For symmetric-teleparallel theories we have instead
\begin{equation}
    \delta\Lambda^\mu{}_\nu=\nabla_\nu\delta\rchi^\mu\to\nabla_\nu\delta\rchi^\mu+\nabla_\nu\xi^\mu\,,
\end{equation}
so the gauge transformation becomes simply
\begin{equation}
    \delta\rchi^\mu\to\delta\rchi^\mu+\xi^\mu\,.
\end{equation}
We can thus always use coordinate transformation to set the affine connection perturbations to zero in symmetric teleparallel theories,
\begin{equation}\label{eq:PertCoincidentGauge}
    \xi^\mu=-\delta\rchi^\mu\,\Rightarrow\,\delta\rchi^\mu\to0\,.
\end{equation}
Note that this allows for a gauge reminiscent of the coincidence gauge, which states that for any given teleparallel and torsion-free connection one may always find a coordinate transformation that makes the connection vanish identically. We have thus found a perturbative version of the coincident gauge with \eqref{eq:PertCoincidentGauge}, since we can always make the connection perturbations vanish by choosing $\xi^\mu=-\delta\rchi^\mu$. This \textit{perturbative coincident gauge} choice fixes the gauge completely.

For the matter part the scalar field perturbation $\delta\Xi$ transforms as
\begin{equation}
    \delta\Xi\to\delta\Xi+\Lie_\xi\Xi=\delta\Xi+\xi^\mu\partial_\mu\Xi\,.
\end{equation}
This completes the way how our involved fields, the metric, the connection, and the matter field, transform under gauge transformations.

\subsection{Scalar-Vector-Tensor splittings}
The cosmological background spacetime we have chosen enjoys spatial homogeneity and isotropy, which foliate the spacetime into a disjoint union of maximally symmetric spatial slices. On each slice one may apply any coordinate transformation respecting these symmetries will leave the metric and teleparallel connection invariant; in the spatially flat case $SO(3)$ rotations are such an example. In the context of cosmological perturbation theory these coordinate transformation allows us to decompose perturbations into irreducable scalar, vector, and tensor modes. Each type of modes evolves independently, and the irreducible representation ensures that the perturbations remain distinct and do not mix with each other as the universe evolves at the level of linear pertubration theory. This mathematical tool simplifies the analysis of perturbations and provides valuable insights into the behavior of different modes in the cosmic evolution.

Let us now discuss the scalar-vector-tensor splittings of the metric and affine connection perturbations. Recall that in a cosmological setup one can split the metric perturbations into how they behave under spatial coordinate transformations, namely scalar, vector, and tensor perturbations \cite{Mukhanov:2005sc}. We now fix the manifold to describe a four dimensional spatially homogeneous and isotropic expanding universe, i.e.
\begin{equation}\label{eq:BackgroundMetric}
    ds^2=-N(\eta)^2d\eta^2+a(\eta)^2\mathfrak{g}_{ij}(x^k)dx^idx^j\,
\end{equation}
at the background level, where $a$ is the scale factor, $N$ is the lapse, and the spatial metric $\mathfrak{g}_{ij}$ is homogeneous and isotropic. The expansion is determined by the conformal Hubble factor $\HH=a'/a$. We can then use the metric perturbations
\begin{align}
    \delta g_{00}&=-2N^2\phi\,,\\
    \delta g_{0i}&=a^2(\d_i j+S_i)\,,\\
    \delta g_{ij}&=2a^2\left(-\g_{ij}\psi-\frac13\g_{ij}\Delta\sigma +\d_i\d_j\sigma+\d_{(i}F_{j)}+h_{ij}\right)\,,
\end{align}
where $\d_i$ denotes the covariant derivative associated to $\g_{ij}$, and $i,j,...$ are spatial indices only. $\Delta=\g^{ij}\d_i\d_j$ with $\g^{ij}$ the inverse of $\g_{ij}$, i.e. $\g^{ij}\g_{jk}=\delta^i_k$, and spatial indices are raised and lowered in the following by $\g_{ij}$. The metric perturbations encompass four scalar modes $\phi,j,\psi,\sigma$, four vector modes $S^i,F^i$ with $\d_iS^i=\d_iF^i=0$, and two tensor modes $h_{ij}$ with $h_{ij}=h_{ji}$, $\d_ih^{ij}=0$, and $\g^{ij}h_{ij}=0$. Splitting similarly the infinitesimal coordinate transformation $\xi^\mu$ into $\xi^0$ and $\xi^i=\d^i\xi+\xi^i_\perp$, with $\d_i\xi^i_{\perp}=0$, the metric perturbations change under gauge transformations as follows,
\begin{align}
    \phi&\to\phi+{\xi^0}'+\ln(N)'\xi^0\,,\\
    j&\to j+\xi'-\frac{N^2}{a^2}\xi^0\,,\\
    \psi&\to\psi-\HH\xi^0-\frac{\Delta}{3}\xi\,,\\
    \sigma&\to\sigma+\xi\,,\\
    S^i&\to S^i+{\xi_{\perp}^i}'\,,\\
    F^i&\to F^i+\xi_{\perp}^i\,,\\
    h_{ij}&\to h_{ij}\,,
\end{align}
where primes are derivatives w.r.t. the conformal time $\eta$ unless stated otherwise. Similarly, we can split the general teleparallel affine connection perturbations as $\delta\Lambda_{\mu\nu}=\delta\Lambda_{(\mu\nu)}+\delta\Lambda_{[\mu\nu]}$, and define its symmetric part analogous to the metric by
\begin{align}
    \delta\Lambda_{(00)}&=-2N^2\phi_c\,,\\
    \delta\Lambda_{(0i)}&=a^2(\d_ij_{c}+S_{ci})\,,\\
    \delta\Lambda_{(ij)}&=2a^2\left(-\g_{ij}\psi_c-\frac13\g_{ij}\Delta\sigma_c +\d_i\d_j\sigma_{c}+\d_{(i}F_{cj)}+h_{cij}\right)\,.
\end{align}
The antisymmetric part we can write similarly to the electromagnetic field strength tensor as
\begin{align}
    \delta\Lambda_{[0i]}&=\frac{a^2}{2}\left(\d_ie+E_i\right)\,,\\
    \delta\Lambda_{[ij]}&=a^2\left(\epsilon_{ijk}\d^kb+\d_{[i}B_{j]}\right)\,,
\end{align}
where $\epsilon_{ijk}=\sqrt{\g}\varepsilon_{ijk}$ with $\varepsilon_{ijk}=0,\pm1$ the totally antisymmetric tensor\footnote{We fix $\varepsilon_{123}=1$.}, and $\g=\det(\g_{ij})$. We then have six scalar modes $\phi_c,j_c,\psi_c,\sigma_c,e,b$, eight vector modes $S_c^i,F_c^i,E^i,B^i$ with $\d_iS^i_c=\d_iF^i_c=\d_iE^i=\d_iB^i=0$, and two tensor modes $h_{cij}$ with $\d_ih_c^{ij}=\g^{ij}h_{cij}=0$. 

Since the gauge transformations depend on the background teleparallel connection though $M^\alpha{}_{\mu\nu}$ we need to specify its general form for cosmological backgrounds. One has by the required homogeneity and isotropy the only non-zero components 
\begin{align}
    M_{000}&=M_1\,,\,M_{0ij}=M_2\g_{ij}\,,\,M_{i0j}=M_3\g_{ij}\,,\,M_{ij0}=M_4\g_{ij}\,,\,M_{ijk}=M_5\epsilon_{ijk}
\end{align}
in form similar to the Levi-Cevita connection of the background \eqref{eq:BackgroundMetric}
\begin{equation}
    \LCBig{0}{00}=\frac{N'}{N}\,,\,\LCBig{0}{ij}=\frac{a^2}{N^2}\HH\g_{ij}\,,\,\LCBig{i}{0j}=\LCBig{i}{j0}=\HH\delta^i_j\,.
\end{equation}
The last non-zero components $\LCBig{k}{ij}$ are just the connection components $\gamma^k{}_{ij}$ associated to $\d_i$,
\begin{equation}
    \LCBig{k}{ij}=\frac12\g^{kl}\left(2\partial_{(i}\g_{j)l}-\partial_l\g_{ij}\right)=\gamma^k{}_{ij}\,.
\end{equation}
$M_1,...,M_5$ are arbitrary functions of conformal time that determine the background connection\footnote{Since $M^k{}_{ij}$ is a tensor under spatial coordinate transformation but $\gamma^k{}_{ij}$ is a connection $M^k{}_{ij}$ cannot have a component proportional to $\gamma^k{}_{ij}$.}. The component $M_5$ is here an axial scalar w.r.t. spatial coordinate transformation, while the rest are polar scalars. We can then derive the gauge transformations for the symmetric variables $\delta\Lambda_{(\mu\nu)}$,
\begin{align}
    \phi_c&\to\phi_c+\frac12\left({\xi^0}'+\left(\ln(N)'-\frac{M_1}{N^2}\right)\xi^0\right)\,,\\
    j_c&\to j_c+\frac12\left(\xi'-\frac{N^2}{a^2}\xi^0+(M_2+M_4)\xi\right)\,,\\
    \psi_c&\to\psi_c-\frac12(\HH+M_3)\xi^0-\frac{\Delta}{6}\xi\,,\\
    \sigma_c&\to\sigma_c+\frac12\xi\,,\\
    S_{c}^i&\to S_c^i+\frac12\left({\xi_\perp^i}'+(M_2+M_4)\xi_\perp^i\right)\,,\\
    F_c^i&\to F_c^i+\frac12\xi_\perp^i\,,\\
    h_{cij}&\to h_{cij}\,,
\end{align}
and the antisymmetric variables of the connection $\delta\Lambda_{[\mu\nu]}$,
\begin{align}
    e&\to e+\xi'+2(\HH+M_2-M_4)\xi-\frac{N^2}{a^2}\xi^0\,,\\
    b&\to b-\frac{M_5}{a^2}\xi\,,\\
    E^i&\to E^i+{\xi_\perp^i}'+2(\HH+M_2-M_4)\xi_\perp^i\,,\\
    B^i&\to B^i-\xi_\perp^i+M_5\tilde\xi_\perp^i\,,
\end{align}
where we defined $\tilde\xi^i_\perp$ by
\begin{equation}
    \d_{[i}\tilde\xi_{\perp j]}=-\epsilon_{ijk}\xi_\perp^k\,.
\end{equation}
In summary, there are a total of 16 perturbation variables in the connection. Among these, one scalar, denoted as $b$, exhibits an axial nature, necessarily decoupling from the other scalar perturbations at first-order perturbation theory unless the background affine geometry contains an axial component in the connection $\Gamma^i{}_{ij}$, i.e. $\epsilon^{ijk}\Gamma_{ijk}\neq0$. Additionally, it is essential to note that $M_5$ must also be axial. Consequently, if the axial component of the connection vanishes, leading to $M_5=0$, the scalar $b$ becomes gauge invariant.

In the metric-teleparallel case we have $2\delta\Lambda_{(\mu\nu)}=\delta g_{\mu\nu}$, so
\begin{align}
    \phi_c=\frac12\phi\,,\,\psi_c=\frac12\psi\,,\,j_c=\frac12j\,,\,\sigma_c=\frac12\sigma\,,\,S^i_c=\frac12S^i\,,\,F^i_c=\frac12F^i\,,\,h_{cij}=\frac12h_{ij}\,.
\end{align}
Only $e,b,E^i,B^i$ remain as independent connection perturbation variables. Their gauge transformations remain of the same form. The connection has two scalar and four vector perturbations, one scalar being axial. There are no tensor perturbations.

For symmetric-teleparallism we have $\delta\rchi^\mu$ as the perturbation variables, which we can split as
\begin{align}
    \delta\rchi^0&=\Chi\,,\\
    \delta\rchi^i&=\d^i\Theta+\Upsilon^i\,,
\end{align}
with $\d_i\Upsilon^i=0$. Their gauge transformations are independent of any background variables,
\begin{align}
    \Chi&\to\Chi+\xi^0\,,\\
    \Theta&\to\Theta+\xi\,,\\
    \Upsilon^i&\to\Upsilon^i+\xi^i_\perp\,.
\end{align}
We have now two scalar and two vector perturbations, and no tensor perturbations.

A surprising result - given that the connection $\Gamma^\alpha{}_{\mu\nu}$ has initially three indices - is that only in the most general teleparallel theory we have tensor perturbations in the connection, while metric or symmetric teleparallel theories have none. The total amount of perturbation variables for each teleparallel theory are summarized in table \ref{tab:perts}. There are no spin-3 perturbation variables with three spatial indices in teleparallel perturbation theory.

 In cosmological perturbation theory, it is useful to introduce gauge invariant perturbation variables. These are quantities that remain unchanged under gauge transformations. In cosmology, these variables are used to describe perturbations in a way that is independent of the chosen coordinate system or gauge. By using gauge invariant perturbation variables, one can effectively analyze and compare perturbations in different cosmological models, ensuring that the physical information is preserved regardless of the gauge choice.
In the scalar sector we can construct the Bardeen potentials $\Phi,\Psi$ as gauge invariant quantities
\begin{align}
    \Phi&=\phi+\frac{a^2}{N^2}\left((2\HH-\ln(N)')(j-\sigma')+(j'-\sigma'')\right)\,\\
    \Psi&=\psi-\frac{a^2}{N^2}\HH(j-\sigma')+\frac{\Delta}{3}\sigma\,.
\end{align}
In the vector sector we have the gauge invariant variables
\begin{equation}
    V^i={S^i}-{F^i}'\,.
\end{equation}
For simplicity we call all $\Phi,\Psi,V^i$ the Bardeen potentials. The tensor perturbations $h_{ij}$ are already gauge invariant. Similarly one can construct gauge invariant variables for the connection ones, but this will not be needed.

\begin{table}
    \centering
    \begin{tabular}{|c|c|c|c|c|}
        \hline
         & Scalars (\#) & Vectors (\#) & Tensors (\#) & Axial scalars (\#) \\
         \hline
          GT & $\phi,j,\psi,\sigma,\phi_c,j_c,\psi_c,\sigma_c,e$ (9) & $S^i,F^i,S^i_c,F^i_c,E^i,B^i$ (12) & $h_{ij},h_{cij}$ (4) & $b$ (1) \\
          \hline
          MT & $\phi,j,\psi,\sigma,e$ (5) & $S^i,F^i,E^i,B^i$ (8) & $h_{ij}$ (2)& $b$ (1)  \\
          \hline
          ST & $\phi,j,\psi,\sigma,\Chi,\Theta$ (6) & $S^i,F^i,\Upsilon^i$ (6) & $h_{ij}$ (2)& -/- \\\hline
    \end{tabular}
    \caption{Perturbation variables in general teleparallel (GT), metric teleparallel (MP), and symmetric teleparallel (ST) theories. \# is the number of perturbation variables in each sector, 26 in total for GT, 16 for MT, and 14 for ST.}
    \label{tab:perts}
\end{table}
In the matter sector we demand from homogeneity that the background scalar field $\Xi$ depends only on time, $\Xi=\Xi(\eta)$. Its perturbation transforms under a gauge transformation as
\begin{equation}\label{eq:XiGaugeTransformation}
    \delta\Xi\to\delta\Xi+\xi^0\Xi'\,,
\end{equation}
such that the combination
\begin{equation}
    \overline{\delta\Xi}=\delta\Xi+\frac{a^2}{N^2}(j-\sigma')\Xi'
\end{equation}
is gauge invariant.

In the following we will focus on both spatially flat and curved universes, i.e. we consider the flat spatial metric
\begin{equation}\label{eq:flatcosmo}
    \g_{ij}=\delta_{ij}\,,\d_i=\partial_i\,,
\end{equation}
in cartesian coordinates $x^i=(x,y,z)$, as well as the curved one
\begin{equation}
    \g_{ij}=\text{diag}\left(\frac{1}{1-u^2r^2},r^2,r^2\sin(\theta)^2\right)
\end{equation}
with spherical coordinates $x^i=(r,\theta,\varphi)$. In the latter case $u$ is the spatial curvature parameter, with $u^2\in\mathbb{R}$. $u^2>0$ corresponds to closed universes and $u^2<0$ to open ones. By setting $u=0$ we can return to the spatially flat case \eqref{eq:flatcosmo} by going from spherical coordinates $(r,\theta,\varphi)$ back to cartesian coordinates $(x,y,z)$ by the usual relation
\begin{equation}
    x=r\cos(\varphi)\sin(\theta)\,,\,y=r\sin(\varphi)\sin(\theta)\,,\,z=r\cos(\theta)\,,
\end{equation}
which leads to the metric becoming flat, $\g_{ij}\to \delta_{ij}$. We also fix\footnote{This can always be achieved by a suitable transformation of the time coordinate.} the lapse to be equal to the scale factor,
\begin{equation}
    N=a\,.
\end{equation}

Lastly, as usual with spatially homogeneous backgrounds, we decompose the perturbations into spherical harmonics - or plane waves in the spatially flat case - which evolve independently thanks to the spatial homogeneity and isotropy. We thus simply set $\Delta=\d^i\d_i\to-k^2$ whenever the Laplacian $\Delta$ acts on a perturbation variable. The ranges of $k^2$ depend here on the background and on which types of modes they act on, see \cite{Bahamonde:2022ohm,1986ApJ}.

\section{General teleparallel theory}

In this section, we shall derive the comprehensive form of the action for general teleparallel perturbations. From this action, the ensuing step will be to derive the equations of motion governing these perturbations. We start with an action of the following structure
\begin{equation}\label{eq:TPAction}
    S=S_{TP}+S_\lambda+S_M=\int\dd^dx\left(\sqrt{-g}\Lie(g_{\mu\nu},M^\alpha{}_{\mu\nu})+\lambda_\alpha{}^{\beta\mu\nu}R^\alpha{}_{\beta\mu\nu}\right)+S_M\,,
\end{equation}
where $S_\lambda$ is the action containing the Lagrange multiplier part, and $\Lie$ is an arbitrary function.

Let us revisit the procedure for deriving the background equations of motion from \eqref{eq:TPAction}. This review will prove instrumental in subsequent steps when we derive the action for perturbations. By varying the action with respect to the metric, we obtain the equations of motion, which can be schematically expressed in the following form
\begin{equation}
    \E_{\mu\nu}=T_{\mu\nu}\,,
\end{equation}
with
\begin{equation}
    \E_{\mu\nu}=\frac{1}{\sqrt{-g}}\frac{\delta S_{TP}}{\delta g^{\mu\nu}}
\end{equation}
and the energy-momentum tensor (EMT)
\begin{equation}
    T_{\mu\nu}=-\frac{1}{\sqrt{-g}}\frac{\delta S_M}{\delta g^{\mu\nu}}\,.
\end{equation}
Varying \eqref{eq:TPAction} w.r.t. to the connection leads to
\begin{equation}\label{eq:TPGammaVariation}
    \sqrt{-g}(Y_\alpha{}^{\mu\nu}-\HH_\alpha{}^{\mu\nu})-2(\nabla_\rho+T_\rho)\lambda_\alpha{}^{\nu\rho\mu}+T^\mu{}_{\rho\sigma}\lambda_\alpha{}^{\nu\rho\sigma}=0\,,
\end{equation}
with
\begin{equation}
    Y_\alpha{}^{\mu\nu}=\frac{1}{\sqrt{-g}}\frac{\delta S_{TP}}{\delta\Gamma^\alpha{}_{\mu\nu}}\,,
\end{equation}
and the hypermomentum tensor
\begin{equation}
    \HH_\alpha{}^{\mu\nu}=-\frac{1}{\sqrt{-g}}\frac{\delta S_{M}}{\delta\Gamma^\alpha{}_{\mu\nu}}\,.
\end{equation}
\eqref{eq:TPGammaVariation} still contains the Lagrange multiplier, but we can get rid of it by taking the divergence $\nabla_\mu$ of \eqref{eq:TPGammaVariation} and using the contracted version of the  Bianchi identity \eqref{eq:BianchiI}, leading to
\begin{equation}
    \C_\alpha{}^\nu=(\nabla_\mu+T_\mu)\left(\sqrt{-g}(Y_\alpha{}^{\mu\nu}-\HH_\alpha{}^{\mu\nu})\right)=0\,,
\end{equation}
We call this the connection equation of motion. In addition the general covariance of the action, see \cite{Heisenberg:2022mbo}, leads to\footnote{The Lagrange multiplier part $S_\lambda$ does not contribute to the Bianchi identity.} the Bianchi identity
\begin{equation}\label{eq:TPBianchi}
    2\sqrt{-g}\D^\mu(\E_{\mu\nu}-T_{\mu\nu})+(\nabla_\mu+T_\mu)\C_\nu{}^\mu+T^\alpha{}_{\mu\nu}\C_\alpha{}^\mu=0\,.
\end{equation}

\subsection{Perturbation action}
The equations of motion for the perturbation can be obtained directly by plugging the perturbations of the metric and connection in the action \eqref{eq:TPAction} and expanding it to second order; the quadratic action of the perturbations we call $\delta^{(2)}S$. Varying w.r.t. to a perturbation variable $v$ then leads to their equation of motion
\begin{equation}
    \E_v=\frac{1}{2\sqrt{-g}}\frac{\delta(\delta^{(2)}S)}{\delta v}=0\,.
\end{equation}

The second order expansion of the teleparallel and matter parts, $S_{TP}$ and $S_M$, is straightforward, but there is also a contribution coming from the Lagrange multiplier piece $S_\lambda$ due to the second order term appearing in the Riemann tensor,
\begin{equation}
    \delta^{(2)}S_\lambda=\int\dd^dx\,\lambda_\alpha{}^{\beta\mu\nu}\delta^{(2)}R^\alpha{}_{\beta\mu\nu}=\int\dd^dx\,\lambda_\alpha{}^{\beta\mu\nu}2\delta\Gamma^\alpha{}_{[\mu|\lambda|}\delta\Gamma^\lambda{}_{\nu]\beta}\,,
\end{equation}
where $\delta^{(2)}$ refers to the second order expansion. After plugging in the connection perturbations and integrating by parts we can use \eqref{eq:TPGammaVariation} to eliminate the Lagrange multiplier in favour of $Y_\alpha{}^{\mu\nu}$ and hypermomentum, which do not depend on Lagrange multipliers,
\begin{align}
    \delta^{(2)}S_\lambda&=\int\dd^dx\,2\lambda_\alpha{}^{\beta\mu\nu}\nabla_\mu\delta\Lambda^\alpha{}_{\lambda}\nabla_\nu\delta\Lambda^\lambda{}_\beta=-\int\dd^dx\,2(\nabla_\mu+T_\mu)\lambda_\alpha{}^{\beta\mu\nu}\nabla_\nu\delta\Lambda^\lambda{}_\beta\delta\Lambda^\alpha{}_\lambda-\nonumber\\
    &-\lambda_\alpha{}^{\beta\mu\nu}2\nabla_{[\mu}\nabla_{\nu]}\delta\Lambda^\lambda{}_{\beta}\delta\Lambda^\alpha{}_\lambda=-\int\dd^dx\,\sqrt{-g}(Y_\alpha{}^{\nu\beta}-\HH_\alpha{}^{\nu\beta})\nabla_\nu\delta\Lambda^\lambda{}_\beta\delta\Lambda^\alpha{}_\lambda\,.
\end{align}
This term will thus contribute friction and mass terms to the connection perturbations. Hence, we have the total action of the 26 perturbation variables 
\begin{equation}
(\phi,j,\psi,\sigma,S^i,F^i,h_{ij},\phi_c,j_c,\psi_c,\sigma_c,S^i_c,F^i_c,h_{cij},e,E^i,b,B^i)\,,
\end{equation}
 and the matter perturbation $\delta\Xi$ of the form
\begin{equation}\label{eq:TPPerturbationAction}
    \delta^{(2)}S=\delta^{(2)}S_{TP}-\int\dd^dx\,\sqrt{-g}(Y_\alpha{}^{\nu\beta}-\HH_\alpha{}^{\nu\beta})\nabla_\nu\delta\Lambda^\lambda{}_\beta\delta\Lambda^\alpha{}_\lambda+\delta^{(2)}S_M\,.
\end{equation}

\subsection{Flat $f(G)$ cosmology}
A straightforward non-GR equivalent example involves setting $\Lie=f(G)$, where $f$ represents a smooth function. In the context of cosmological symmetries, three potential connections are compatible in the spatially flat case, as discussed in \cite{Heisenberg:2022mbo}.
Connection 1 has the form $\Gamma^0{}_{00}=\HH-\Lie$, $\Gamma^i{}_{0j}=(\HH+\K)\delta^i_j$, and all other components zero. Connection 2 has only $\Gamma^0{}_{00}=\HH+\K-\Lie'/\Lie$, $\Gamma^0{}_{ij}=-\Lie\delta_{ij}$, and $\Gamma^i{}_{0j}=(\HH+\K)\delta^i_j$ non-zero. Connection 3 has only $\Gamma^0{}_{00}=\HH+\K+\Lie'/\Lie$, $\Gamma^i{}_{j0}=\Lie\delta^i_j$, and $\Gamma^i{}_{0j}=(\HH+\K)\delta^i_j$ non-vanishing. $\K,\Lie$ are two free functions of conformal time. Note however that $\K$ never appears in the background equations of motion.

For our choice of matter we have vanishing hypermomentum, but the scalar field contributes to the EMT. The background is however still rather simple \cite{Heisenberg:2022mbo}. For connection 1 the connection equation of motion is trivial, and $\Lie$ does not appear in the background. The background equations of motion are
\begin{align}
    G&=\frac{6\HH^2}{a^2}\,,\label{eq:BGCosmo0}\\
    f(G)&=2Gf'(G)-\rho\,,\label{eq:BGCosmo1}\\
    \HH'&=\HH^2-\frac{a^2{\Xi'}^2}{4a^2f'(G)+48\HH^2f''(G)}\label{eq:BGCosmo2}\,,
\end{align}
where the energy density $\rho$ of the scalar field is
\begin{equation}
    \rho=\frac{{\Xi'}^2}{2a^2}+\mathcal{V}=\frac{2}{a^2}T_{00}\,.
\end{equation}
Note that primes on $f$ are derivatives w.r.t. its argument and not conformal time, i.e. $f'(G)=df/dG$, and similarly for $f(\T)$ and $f(\Q)$ cosmology below. In addition we have the scalar equation of motion
\begin{equation}
    \Xi''+2\HH\Xi'+a^2\mathcal{V}'=0\,.
\end{equation}
For connections 2 and 3, by the connection equations of motion the connection is such that $G=G_0$ is constant, and this in turn fixes the metric equations of motion to have the same form as in GR with an effective cosmological constant determined by $f$ and $G_0$, i.e. we have now the gravitational equations of motion
\begin{align}
    f(G_0)&=\left(G_0+\frac{6\HH^2}{a^2}\right)f'(G_0)-\rho\,,\\
    \HH'&=\HH^2-\frac{{\Xi'}^2}{4f'(G_0)}\,.
\end{align}
The condition $G=G_0$ is fulfilled by choosing an appropriate $\Lie$, e.g. for connection 2
\begin{equation}\label{eq:Lsol2}
    \Lie'=-2\HH^2+\frac{G_0}{3}a^2-2\HH\Lie\,,
\end{equation}
and for connection 3
\begin{equation}\label{eq:Lsol3}
    \Lie'=2\HH^2-\frac{G_0}{3}a^2-2\HH\Lie\,.
\end{equation}
$\Lie$ is then given by an integral - in particular in $\Lie$ there appears an undetermined constant of integration - so instead we just replace $\Lie'$ in the equations of motion by \eqref{eq:Lsol2}, \eqref{eq:Lsol3}.

The reason for including the scalar field $\Xi$ now becomes clear. In its absence $\Xi=0$, or rather $\Xi'=0$, the condition $\HH'=\HH^2$ from the equations of motion fixes
\begin{equation}
    a(\eta)=-\frac{1}{H_0}\frac{1}{\eta-\eta_0}\,,
\end{equation}
with constants of integration $H_0,\eta_0$, leading to the Hubble function $H=\HH/a=H_0=$ const. The spacetime is thus fixed to de Sitter space only, which makes this type of theory rather simple in vacuum. In particular we have then for connection 1 $G=6\HH^2/a^2=6H^2=6H_0^2=\mathcal{R}/2$ constant as well. We keep $\Xi$ to make the background less trivial, but we can always choose $\Xi=0$ to return to de Sitter.

We analyze the perturbations for each connection separately. Since none of the three background connections in the spatially flat case have an axial component, i.e. $\epsilon^{ijk}\Gamma_{ijk}=0$, the axial scalar $b$ necessairly decouples from the remaining polar perturbations.

\subsubsection{Connection 1}

The axial scalar $b$ is here absent, with $\E_b=0$. We solve $\E_\phi=0$ for $\phi$, $\E_j=0$ for $j$, and $\E_e=0$ for $e$. The intermediate solutions are very complicated and we do not report them here. Afterwards, we solve $\E_{j_c}=0$ for $\sigma_c$, and $\E_{\sigma_c}=0$ for $j_c$, and make the replacement
\begin{equation}
\delta\Xi\to\delta\Xi+\frac{\Xi'}{3\HH}(k^2\sigma-3\psi)\,,
\end{equation}
upon which $\E_\psi=\frac{\Xi'}{\HH}\E_{\delta\Xi}$ and $\E_\sigma=-\frac{k^2\Xi'}{3\HH}\E_{\delta\Xi}$, and $\E_{\delta\Xi}$ only contains $\delta\Xi$,
\begin{align}
    \E_{\delta\Xi}&=\frac{1}{2a^2}\left[\delta\Xi''+2\HH\delta\Xi'+\left(k^2+a^2\mathcal{V}''+\frac{\Xi'}{f'}\left(\frac{a^2\mathcal{V'}}{\HH}+\frac{3\Xi'}{2}-\frac{{\Xi'}^3}{8\HH f'}\right)\right)\delta\Xi\right]\,.\label{eq:fGXiEOM}
\end{align}
All other equations of motion are thus fulfilled, and we have integrated out all scalar degrees of freedom except $\delta\Xi$. The solutions read as follows
\begin{dmath}
    {\phi=\ }\frac{1}{12 \mathcal{H}^2 f' \left(a^2 f'+12 \mathcal{H}^2 f''\right)}\Big(a^2 f' \left(\Xi '\right)^2 \left(k^2 \sigma -3 \psi \right)+a^2 \mathcal{H} f' \left(3 \delta \Xi  \Xi '+4 f' \left(k^2 \sigma '-3 \psi '\right)\right)+12 \mathcal{H}^3 f'' \left(3 \delta \Xi  \Xi '+4 f' \left(k^2 \sigma '-3 \psi '\right)\right)\Big)\,,\\
    {j=\ }\frac{\frac{a^2 \delta \Xi  \mathcal{V}'(\Xi )+\delta \Xi ' \Xi '+4 k^2 \psi  f'}{4 k^2 f'}-\frac{k^2 \sigma }{3}}{\mathcal{H}}+\frac{3 \delta \Xi  \Xi '}{4 k^2 f'}-\frac{\delta \Xi  \left(\Xi '\right)^3}{16 k^2 \mathcal{H}^2 \left(f'\right)^2}+\sigma '\,,\\
    {e=\ }\frac{\frac{a^2 \delta \Xi  \mathcal{V}'(\Xi )+\delta \Xi ' \Xi '+4 k^2 \psi  f'}{4 k^2 f'}-\frac{k^2 \sigma }{3}}{\mathcal{H}}-\frac{3 \delta \Xi  \Xi '}{4 k^2 f'}-\frac{\delta \Xi  \left(\Xi '\right)^3}{16 k^2 \mathcal{H}^2 \left(f'\right)^2}-\sigma '\,,\\
    {\sigma_c=\ }\frac{1}{16 k^2 \mathcal{H}^2 f' \left(a^2 f'+12 \mathcal{H}^2 f''\right)}\Big(4 a^2 \mathcal{H}^2 \left(f'\right)^2 \left(-6 \psi _c-6 \phi _c+2 k^2 \sigma +3 \psi \right)+a^2 f' \left(\Xi '\right)^2 \left(k^2 \sigma -3 \psi \right)+a^2 \mathcal{H} f' \left(3 \delta \Xi  \Xi '+4 k^2 f' \sigma '-8 \mathcal{L} f' \left(k^2 \sigma -3 \psi \right)-12 f' \psi '\right)+48 \mathcal{H}^4 f' f'' \left(-6 \psi _c-6 \phi _c+2 k^2 \sigma +3 \psi \right)-12 \mathcal{H}^3 f'' \left(-3 \delta \Xi  \Xi '-4 k^2 f' \sigma '+8 \mathcal{L} f' \left(k^2 \sigma -3 \psi \right)+12 f' \psi '\right)\Big)\,,\\
    {j_c=\ }\frac{1}{96 k^2 \mathcal{H}^2 \left(f'\right)^2}\Big(4 \mathcal{H} f' \left(3 \left(a^2 \delta \Xi  \mathcal{V}'(\Xi )+\delta \Xi ' \Xi '+4 k^2 \psi  f'\right)-4 k^4 \sigma  f'\right)-3 \delta \Xi  \left(\Xi '\right)^3+12 \mathcal{H}^2 f' \left(3 \delta \Xi  \Xi '+4 k^2 f' \sigma '\right)\,,
\end{dmath}
such that the Bardeen potentials become
\begin{align}
    \Phi&=\Psi+f''\frac{9\HH(\HH^2-\HH')\Xi'}{k^2a^2f'}\delta\Xi\,,\\
    \Psi&=\frac{1}{16k^2\HH {f'}^2}\left[-f'\left(\HH a^2\mathcal{V}'\delta\Xi+\Xi'((2\HH^2+\HH')\delta\Xi+\HH\delta\Xi')\right)+2Gf''(\HH^2-\HH')\Xi'\delta\Xi\right]\,.
\end{align}
We see that $\Phi$ and $\Psi$ are now determined completely by the perturbations of the scalar field, so they vanish in its absence, $\delta\Xi=0$. Moreover, we find that unlike in GR\footnote{In GR the equality of the two Bardeen potentials, $\Phi=\Psi$, follows for any matter whose perturbations fulfill $\delta T^i_j=\propto\delta^i_j$, with $\delta T^\mu_\nu$ the perturbations of the EMT. The EMT of the scalar field $\Xi$ fulfills this condition, so in GR we would end up with $\Phi=\Psi$.} we have $\Phi\neq\Psi$ as long as $f''\neq0$ and we are not in de Sitter space, $\HH'\neq\HH^2$. At this point we must note that while the perturbation solutions above actually depend on the undetermined background function $\Lie$, the physical perturbations $\Phi,\Psi$ and $\E_{\delta\Xi}$ are independent of it.

Since the scalar matter field is always dynamical due to its own equation of motion we thus find that no scalar gravitational degrees of freedom remain, even in general backgrounds that are not de Sitter.

In the vector sector we solve $\E_{S^i_c}=\E_{F^ic}=\E_{E^i}=\E_{B^i}=0$ for $S^i_c,F^i_c,E^i,B^i$,
\begin{align}
    F^i_c&=\frac{F^i}{2}\,,\\
    S^i_c&=\frac{S^i}{2}\,,\\
    B^i&=\frac{F^i}{2}\,,\\
    E^i&=-S^i\,,
\end{align}
such that now $\E_{S^i}=0$ are then solved by $S^i={F^i}'$. $\E_{F^i}$ are now vanishing and we have no vector degrees of freedom and $V^i=0$.

In the tensor sector we have $h_{cij}$ absent, with $\E_{h_{cij}}=0$ identically, but we do find more interesting equations of motion for the gravitational wave modes $h_{ij}$,
\begin{equation}
    \frac{f'}{2a^2}\left(h_{ij}''+(2\HH+\ln(f')') h_{ij}'+k^2 h_{ij}\right)=0\,,
\end{equation}
where $\ln(f')'=\partial_\eta \ln(f'(G))$. In vacuum we are in de Sitter space with constant $G$ making this extra term vanish. In more general backgrounds with $G$ not constant this term can have however interesting consequences; while on long wavelengths $k\to 0$ the constant mode $h_{ij}=$ const still dominates, for short wavelengths we have now the decay
\begin{equation}
    h_{ij}\propto \frac{1}{a\sqrt{f'(G)}}
\end{equation}
instead, leading to an additional enhancement/decay by the $1/\sqrt{f'(G)}$ factor depending on the form of $f$ and the background expansion $G\propto H^2$. We see that for the first connection choice, the $f(G)$ cosmology always suffers from the strong coupling problem, where aside the standard tensor perturbations, the scalar and vector sectors completely disappear. 

\subsubsection{Connection 2}

In the scalar sector neither $b$ or $\sigma_c$ appear, and $\E_b=\E_{\sigma_c}=0$. We solve $\E_j=0$ for $j$,
$\E_{e}=0$ for $\phi$, and $\E_\phi=0$ for $e$,
leading to $\E_{\phi_c}=\E_{j_c}=\E_{\psi_c}=0$. 
We can again set
\begin{equation}
\delta\Xi\to\delta\Xi+\frac{\Xi'}{3\HH}(k^2\sigma-3\psi)\,,
\end{equation}
upon which $\E_\psi=\frac{\Xi'}{\HH}\E_{\delta\Xi}$ and $\E_\sigma=-\frac{k^2\Xi'}{3\HH}\E_{\delta\Xi}$, and $\E_{\delta\Xi}$ has the same form as \eqref{eq:fGXiEOM} with now $G=G_0$. Since $G$ is now constant we also have for the Bardeen potentials
\begin{equation}
    \Phi=\Psi=\frac{1}{16 k^2 \mathcal{H} \left(f'\right)^2}\Big(-4 a^2 \delta \Xi  \mathcal{H} f' \mathcal{V}'(\Xi )+\delta \Xi  \left(\Xi '\right)^3-4 \mathcal{H} f' \Xi ' \left(\delta \Xi '+3 \delta \Xi  \mathcal{H}\right)\Big)\,,
\end{equation}
similar to GR\footnote{GR is in this background recovered for $f'(G_0)=1$ and $f(G_0)=G_0$. The undetermined $G_0$ then drops out of the background equations, and they take the same form as in GR.}.

We thus have no gravitational scalar perturbations propagating. The solutions are
\begin{dmath}
    {j=\ }\frac{1}{48 k^2 \mathcal{H}^2 \left(f'\right)^2}\Big(4 \mathcal{H} f' \left(3 \left(a^2 \delta \Xi  \mathcal{V}'(\Xi )+\delta \Xi ' \Xi '+4 k^2 \psi  f'\right)-4 k^4 \sigma  f'\right)-3 \delta \Xi  \left(\Xi '\right)^3+12 \mathcal{H}^2 f' \left(3 \delta \Xi  \Xi '+4 k^2 f' \sigma '\right)\,,\\
    {\phi=\ }\frac{1}{12 \mathcal{H}^2 f'}\Big(\mathcal{H} \left(3 \delta \Xi  \Xi '+4 f' \left(k^2 \sigma '-3 \psi '\right)\right)+\left(\Xi '\right)^2 \left(k^2 \sigma -3 \psi \right)\Big)\,,\\
    {e=\ }\frac{1}{144 k^2 \mathcal{H}^3 \mathcal{L} \left(f'\right)^2 (2 \mathcal{H}+\mathcal{L})}\Big(-6 \mathcal{H}^3 \left(16 k^2 j_c \left(f'\right)^2 \left(a^2 G_0-3 \mathcal{L}^2\right)+2 a^2 G_0 f' \left(24 \mathcal{L} f' \left(\psi _c+\phi _c-\psi \right)-3 \delta \Xi  \Xi '-4 k^2 f' \sigma '\right)+6 \mathcal{L}^2 f' \left(4 f' \left(6 \left(\psi _c'+\phi _c'\right)+k^2 \sigma '\right)+9 \delta \Xi  \Xi '\right)-3 \delta \Xi  \left(\Xi '\right)^3+24 \mathcal{L} f' \left(\Xi '\right)^2 \left(k^2 \sigma -3 \psi \right)\right)+24 \mathcal{H}^4 f' \left(-3 \left(a^2 \delta \Xi  \mathcal{V}'(\Xi )+\delta \Xi ' \Xi '+4 k^2 \psi  f'+8 k^2 \mathcal{L} f' \sigma '+6 \delta \Xi  \mathcal{L} \Xi '\right)+24 k^2 \mathcal{L} j_c f'+4 k^4 \sigma  f'\right)-4 \mathcal{H}^2 f' \left(a^2 \left(4 k^4 \sigma  G_0 f'-3 \left(4 k^2 G_0 \psi  f'+G_0 \delta \Xi ' \Xi '-9 \delta \Xi  \mathcal{L}^2 \mathcal{V}'(\Xi )\right)\right)-3 a^4 \delta \Xi  G_0 \mathcal{V}'(\Xi )-9 \mathcal{L}^2 \left(3 \delta \Xi ' \Xi '+4 k^2 f' \sigma ''-12 f' \psi ''-6 k^2 \sigma  \left(\Xi '\right)^2+18 \psi  \left(\Xi '\right)^2\right)\right)-3 \mathcal{H} \Xi ' \left(a^2 \left(24 \mathcal{L}^2 f' \left(k^2 \sigma -3 \psi \right) \mathcal{V}'(\Xi )+\delta \Xi  G_0 \left(\Xi '\right)^2\right)-3 \mathcal{L}^2 \Xi ' \left(3 \delta \Xi  \Xi '+8 f' \left(k^2 \sigma '-3 \psi '\right)\right)\right)+72 \mathcal{H}^5 f' \left(8 k^2 j_c f'+24 \mathcal{L} f' \left(\psi _c+\phi _c-\psi \right)-3 \delta \Xi  \Xi '-4 k^2 f' \sigma '\right)+18 \mathcal{L}^2 \left(\Xi '\right)^4 \left(k^2 \sigma -3 \psi \right)\Big)\,.
\end{dmath}

The vector and tensor sectors are as for connection 1 with now $G=G_0$ fixed, so no degrees of freedom in addition to $h_{ij}$ appear. In particular, the equation of motion for the gravitational wave modes $h_{ij}$ is now $f'(G_0)$ times the GR equation, so even the gravitational waves propagate now as in a GR background. Thus, this connection choice also suffers from the strong coupling problem.

\subsubsection{Connection 3}
The strategy for solving the scalar part is the same as for connection 2, but with slightly different solutions,
\begin{dmath}
    {j=\ }\frac{1}{48 k^2 \mathcal{H}^2 \left(f'\right)^2}\Big(4 \mathcal{H} f' \left(3 \left(a^2 \delta \Xi  \mathcal{V}'(\Xi )+\delta \Xi ' \Xi '+4 k^2 \psi  f'\right)-4 k^4 \sigma  f'\right)-3 \delta \Xi  \left(\Xi '\right)^3+12 \mathcal{H}^2 f' \left(3 \delta \Xi  \Xi '+4 k^2 f' \sigma '\right)\Big)\,,\\
    {\phi=\ }\frac{1}{12 \mathcal{H}^2 f'}\Big(\mathcal{H} \left(3 \delta \Xi  \Xi '+4 f' \left(k^2 \sigma '-3 \psi '\right)\right)+\left(\Xi '\right)^2 \left(k^2 \sigma -3 \psi \right)\Big)\,,\\
    {e=\ }\frac{1}{144 k^2 \mathcal{H}^3 \mathcal{L} \left(f'\right)^2 (2 \mathcal{H}-\mathcal{L})}\Big(-6 \mathcal{H}^3 \left(16 k^2 j_c \left(f'\right)^2 \left(a^2 G_0-3 \mathcal{L}^2\right)+8 \mathcal{L} f' \left(4 k^4 \sigma  f'-3 \left(a^2 \left(2 G_0 \psi _c f'+2 G_0 \phi _c f'+\delta \Xi  \mathcal{V}'(\Xi )\right)+\delta \Xi ' \Xi '+4 k^2 \psi  f'\right)\right)-8 a^2 k^2 G_0 \left(f'\right)^2 \sigma '-6 a^2 \delta \Xi  G_0 f' \Xi '+18 \mathcal{L}^2 f' \left(4 f' \left(2 \left(\psi _c'+\phi _c'-2 \psi '\right)+k^2 \sigma '\right)+5 \delta \Xi  \Xi '\right)-3 \delta \Xi  \left(\Xi '\right)^3\right)+24 \mathcal{H}^4 f' \left(-3 \left(a^2 \delta \Xi  \mathcal{V}'(\Xi )+\delta \Xi ' \Xi '+4 k^2 \psi  f'-8 k^2 \mathcal{L} f' \sigma '+24 \mathcal{L} f' \psi '-6 \delta \Xi  \mathcal{L} \Xi '\right)-24 k^2 \mathcal{L} j_c f'+4 k^4 \sigma  f'\right)-4 \mathcal{H}^2 \left(a^2 f' \left(4 k^4 \sigma  G_0 f'-3 \left(4 k^2 G_0 \psi  f'-8 k^2 G_0 \mathcal{L} f' \sigma '+24 G_0 \mathcal{L} f' \psi '+G_0 \delta \Xi ' \Xi '-6 \delta \Xi  G_0 \mathcal{L} \Xi '-15 \delta \Xi  \mathcal{L}^2 \mathcal{V}'(\Xi )\right)\right)-3 a^4 \delta \Xi  G_0 f' \mathcal{V}'(\Xi )+3 \mathcal{L} \left(3 \delta \Xi  \left(\Xi '\right)^3+\mathcal{L} f' \left(-3 \delta \Xi ' \Xi '-12 k^2 f' \sigma ''-8 k^4 \sigma  f'+24 k^2 \psi  f'+36 f' \psi ''+18 k^2 \sigma  \left(\Xi '\right)^2-54 \psi  \left(\Xi '\right)^2\right)\right)\right)-3 \mathcal{H} \Xi ' \left(a^2 \left(8 G_0 \mathcal{L} f' \Xi ' \left(k^2 \sigma -3 \psi \right)+24 \mathcal{L}^2 f' \left(k^2 \sigma -3 \psi \right) \mathcal{V}'(\Xi )+\delta \Xi  G_0 \left(\Xi '\right)^2\right)-3 \mathcal{L}^2 \Xi ' \left(5 \delta \Xi  \Xi '+8 f' \left(k^2 \sigma '-3 \psi '\right)\right)\right)+72 \mathcal{H}^5 f' \left(8 k^2 j_c f'-24 \mathcal{L} f' \left(\psi _c+\phi _c\right)-3 \delta \Xi  \Xi '-4 k^2 f' \sigma '\right)+18 \mathcal{L}^2 \left(\Xi '\right)^4 \left(k^2 \sigma -3 \psi \right)\Big)\,.
\end{dmath}
$\E_{\delta\Xi}$ and $\Phi=\Psi$ are as for connection 2. Again no gravitational scalar perturbations propagate.
The vector and tensor sectors are as for connection 2, so no degrees of freedom in addition to $h_{ij}$ appear.
As it becomes clear from our analysis, all the three connection choices give rise to strong coupling problems on generic cosmological backgrounds.
Unfortunately, in this case the framework of perturbation theory breaks down as the higher order perturbations become strongly coupled. This means that $(G)$ theories do not possess viable
cosmological solutions and the standard perturbative cosmological analysis cannot be applied.

\subsection{Curved $f(G)$ cosmology}
Let us now turn to spatially curved geometries with $\Lie=f(G)$. From \cite{Heisenberg:2022mbo} we find that we have two\footnote{Note that since the background metric depends only on $u^2$ the connection may depend on $\pm u$. Since the sign of $u$ is irrelevant for the metric we only consider the case where we choose the $+u$ sign; the other case may then be obtained from our formulas by setting $u\to-u$. There are no qualitative differences between the two cases.} admissible connections; connection 1 has $\Gamma^0{}_{00}=\HH-\Lie$, $\Gamma^i{}_{0j}=(\HH+\K)\delta^i_j$, and $\Gamma^i{}_{jk}=\gamma^i{}_{jk}+\Delta\Gamma^i{}_{jk}$ with $\Delta\Gamma^r{}_{\theta\varphi}=ur^2\sqrt{1-u^2r^2}\sin(\theta)=-\Delta\Gamma^r{}_{\varphi\theta}$, $\Delta\Gamma^\theta_{\varphi r}=u\sin(\theta)/\sqrt{1-u^2r^2}=-\Delta\Gamma^\theta{}_{r\varphi}$, and $\Delta\Gamma^\varphi{}_{r\theta}=u\csc(\theta)/\sqrt{1-u^2r^2}=-\Delta\Gamma^\varphi{}_{\theta r}$. All other components vanish. Note that this connection now has an axial component,
\begin{equation}
    \epsilon^{ijk}M_{ijk}=\epsilon^{ijk}\Gamma_{ijk}=6u\,,
\end{equation}
non-vanishing for all $u\neq0$. Since the background affine geometry now has an axial component, the axial perturbation $b$ may not decouple from the other scalar perturbations. For this connection the background equations of motion takes the following form for the metric,
\begin{align}
    G&=\frac{6(\HH^2-u^2)}{a^2}\,,\\
    f(G)&=2Gf'(G)-\rho\,,\\
    \HH'&=\HH^2-\frac{a^2{\Xi'}^2}{4a^2f'(G)+48\HH^2f''(G)}+u^2\frac{a^2f'(G)-12\HH^2f''(G)}{a^2f'(G)+12\HH^2f''(G)}\,,
\end{align}
with $\rho$ the same energy density of $\Xi$ as in the flat case. We see that the spatial curvature $u^2$ appears explicitly in $G$. The equation of motion for $\Xi$ has the same form as in the spatially flat case. The connection equation of motion is also identically fulfilled; in particular the free functions $\Lie,\K$ are again undetermined by the background equations of motion.

Another possible background connection is given by $\Gamma^0{}_{00}=\HH+\K+\Lie'/\Lie$, $\Gamma^i{}_{0j}=(\HH+\K)\delta^i_j$, $\Gamma^i{}_{j0}=\Lie\delta^i_j$, and $\Gamma^0{}_{ij}=-\frac{u^2}{\Lie}\g_{ij}$. $\Delta\Gamma^i{}_{jk}$ and all remaining components vanish, as well as the axial component, $\epsilon^{ijk}\Gamma_{ijk}=0$. For this connection the connection equation of motion is non-trivial, and - since for our choice of matter the hypermomentum vanishes - demands \cite{Heisenberg:2022mbo}
\begin{equation}
    G'(u^2+\Lie^2)=0\,.
\end{equation}
We have here now two possibilities for choosing $\Lie$; similar to connections 2 and 3 from the flat case we can choose $\Lie$ such that $G=G_0$ is constant. This is achieved by setting
\begin{equation}
    \Lie'=\frac{\Lie}{3(\Lie^2+u^2)}\left(6\HH^2\Lie-\Lie(G_0a^2+6u^2)+6\HH(u^2-\Lie^2)\right)\,,
\end{equation}
which we call connection 2. The background equations of motion take the GR-like form
\begin{align}
    f(G_0)&=f'(G_0)\frac{6(\HH^2+u^2)}{a^2}+f'(G_0)G_0-\rho\,,\\
    \HH'&=\HH^2+u^2-\frac{{\Xi'}^2}{4f'(G_0)}\,,
\end{align}
as well as the same $\Xi$ equation of motion as in the flat case.

Another possibility\footnote{In the spatially flat case this choice was not possible, as then $u=\Lie=0$, which is incompatible with the background connection that contains $1/\Lie$ terms.} is $\Lie=\pm iu\neq 0$, which is a viable choice only if $u^2<0$ since $\Lie$ should be real. We call this connection 3. Now we have the background equations of motion
\begin{align}
    G&=\frac{6(\HH-iu)^2}{a^2}\,,\\
    f(G)&=\frac{12\HH}{a^2}(\HH-iu)f'(G)-\rho\,,\\
    \HH'&=\frac{a^2f'(G)(\HH^2+u^2)+12i(u+i\HH)^3\HH f''(G)}{a^2f'(G)-12(u+i\HH)^2f''(G)}-\frac{a^2{\Xi'}^2}{4(a^2f'(G)+12(\HH-iu)^2f''(G))}\,.
\end{align}
These equations are again only real if $u^2<0$.

\subsubsection{Connection 1}
Since the connection now contains a non-vanishing axial part we find in the scalar sector that $b$ does appear in the equations of motion of the other scalar perturbations, and
\begin{align}
    \E_b&=\frac{f''(G)k^2u}{4}\left[\frac{16}{a^4}\left(2k^2u\,b-\frac{k^2}{2}(j-2j_c)(\K+\Lie)+6\HH^2\phi+6u^2\psi+\HH(6\psi'-k^2(j-2e))-\right)\right.\\
    &\left.-12\HH\frac{8u^2f'(G)-{\Xi'}^2}{a^4f'(G)+12\HH^2a^2f''(G)}(j+2e)\right]
\end{align}
no longer vanishes. However, this equation is algebraic in $b$, and may be solved for it as long as $u^2f''(G)\neq 0$. Upon plugging this solution into the remaining equations of motion for the scalar perturbations, we can solve them in the same way as in the flat case, namely by solving $\E_{\phi,j,\psi_c,j_c,e}=0$ for $\phi,j,j_c,\psi_c,e$. Shifting again $\delta\Xi\to\delta\Xi-\frac{\Xi'}{\HH}\left(\psi-\frac{k^2}{3}\sigma\right)$ leads to $\psi,\sigma$ dropping out of the remaining non-zero equations of motion, which fulfill $\E_\psi=\frac{\Xi'}{\HH}\E_{\delta\Xi}$ and $\E_\sigma=-\frac{k^2\Xi'}{3\HH}\E_{\delta\Xi}$. The only equation of motion remaining is then $\E_{\delta\Xi}$, which is a second order equation of motion for $\delta\Xi$ only. All appearances of the undetermined functions $\K,\Lie$ have also disappeared from $\E_{\delta\Xi}$, as well as from the Bardeen potentials $\Phi,\Psi$ which depend on $\delta\Xi$ only. The solutions are lengthy and not insightful, so we do not report them here. However, it should be noted that in the flat limit $u\to 0$ we have
\begin{align}
    b&=\frac{1}{u}\frac{(a^2f'(G)+12\HH^2f''(G))(\delta\Xi{\Xi'}^3-4\HH f'(G)(a^2\mathcal{V}'\delta\Xi+\Xi'\delta\Xi'))}{4\HH(a^2f'(G)^2(8k^2f'(G)+3{\Xi'}^2)+3f''(G)({\Xi'}^4+4\HH^2f'(G)(8k^2f'(G)+3{\Xi'}^2)))}\,.
\end{align}
The right hand side depends only on the initial values of $\delta\Xi$ and $\delta\Xi'$, so in general this diverges as $b\propto 1/u$ when going to a spatially flat universe. 

In the vector sector we can solve $\E_{S^i,S^i_c,F^i_c,E^i,B^i}=$, for $S^i,S^i_c,F^i_c,B^i,E^i$,
\begin{align}
    S^i&={F^i}'\,,\\
    S^i_c&=\frac{{F^i}'}{2}\,,\\
    F^i_c&=\frac{F^i}{2}\,,\\
    E^i&=-{F^i}'\,,\\
    B^i&=\frac{(k^2-2u^2)F^i}{2(k^2+2u^2)}\,,
\end{align}
upon which $\E_{F^i}=0$ and $V^i=0$. We have thus integrated out all vector modes.

In the tensor sector we find $h_{cij}$ absent and $\E_{h_{cij}}=0$ identically, and
\begin{align}
    \E_{h_{ij}}=\frac{f'(G)}{2a^2}\left(h''_{ij}+(2\HH+\ln(f'(G))')h_{ij}'+(k^2+2u^2)h_{ij}\right)\,,
\end{align}
so it takes the usual form of a gravitational wave in curved cosmology with the extra term $\propto f''(G)G'$ leading to the additional $1/\sqrt{f'(G)}$ decay of short wavelength modes.

In conclusion we find again no propagating modes except the two tensor ones. However the axial scalar $b$ is no longer absent, and it is determined by $\E_b\propto u$. This is in stark contrast to the spatially flat background where $b$ was left arbitrary. This is again a strong coupling problem since scalar and vector degrees of freedom loose their kinetic term, and only tensor perturbations propagate.

\subsubsection{Connection 2}

Since this connection leads to very GR like background dynamics we find that $\E_b=0$ and $b$ is still absent and left unfixed. The remaining equations of motion are solved similar to the spatially flat case with connections 2 and 3 that also had $G=G_0$ constant. All scalar perturbations except $\delta\Xi$ can be integrated out and are not dynamical; in particular the $u\to 0$ limit does not lead to divergences in the perturbations.
In the vector sector we have $\E_{S^i_c,F^i_c,E^i,B^i}=0$ identically, and after solving $\E_{S^i}=0$ for $S^i={F^i}'$ we find $\E_{F^i}=0$ and $V^i=0$, so all vector modes have been integrated out.
The tensor sector has $h_{cij}$ absent, $\E_{h_{cij}}=0$, and $\E_{h_{ij}}$ proportional the GR equation of motion for gravitational waves.

In conclusion we find that in this very GR like sector we do not find any new degrees of freedom propagating, suffering again from strong coupling problem. The axial sector given by $b$ is as in the flat geometry unfixed.

\subsubsection{Connection 3}

In the scalar sector we find that $b$ decouples from the remaining perturbations since the background connection has no axial component, but $\E_b$ still does not vanish identically,
\begin{equation}
    \E_b=-\frac{2iu\,f'(G)'}{a^2}\,b\,.
\end{equation}
This equation is algebraic in $b$ and has the trivial solution $b=0$. We then solve $\E_{\phi,j,\phi_c,j_c,\sigma_c,e}=0$ for $\phi,j,\psi_c,\sigma_c,j_c,e$. Performing similar manipulations to $\delta\Xi$ one finds similar results as for connection 1, i.e. all scalar perturbations except $\delta\Xi$ can be integrated out. In this case however the solution for $b=0$ is trivially stable for $u\to 0$, even though $\E_b$ vanishes in that flat limit where $b$ is undetermined. In the vector sector we solve the equations as for connection 1, but now with the solutions
\begin{align}
    S^i&={F^i}'\,,\\
    S^i_c&=\frac{{F^i}'}{2}\,,\\
    F^i_c&=\frac{F^i}{2}\,,\\
    E^i&=-2iuF^i-{F^i}'\,,\\
    B^i&=\frac{1}{2}F^i\,,
\end{align}
upon which $\E_{F^i}=0$ and $V^i=0$ again, so all vector modes can be integrated out.
In the tensor sector we find a non-vanishing equation of motion for $h_{cij}$,
\begin{align}
    \E_{h_{cij}}&=\frac{iu f'(G)'}{a^2}(2h_{cij}-h_{ij})\,,
\end{align}
which has the algebraic solution
\begin{equation}
    h_{cij}=\frac12 h_{ij}
\end{equation}
for $\E_{h_{cij}}=0$. We then have the equation for $h_{ij}$,
\begin{equation}
    \E_{h_{ij}}=\frac{f'(G)}{2a^2}\left(h_{ij}''+(2\HH+\ln(f'(G))')h_{ij}'+(k^2+2u^2+iu\ln(f'(G))')h_{ij}\right)\,,
\end{equation}
which has not only the new term $\propto f''(G)G'$ in the friction but also in the mass term, leading to a different non-constant solution in the long wavelength limit $k^2+2u^2\to0$ if $\ln(f'G))'$ is large. This can lead to differences in the evolution of superhorizon modes during periods where $G$ changes rapidly.

For this choice of background connection we find that for $u,G'\neq 0$ the axial scalar field $b$ has to vanish, unlike for the spatially flat geometries where $b$ was undetermined. We also find in the tensor sector that the connection tensor modes $h_{cij}$ are fixed to $h_{ij}/2$ by their equations of motion only if $u,G'\neq 0$, whereas in the spatially flat case $h_{cij}$ was left arbitrary. In addition, we see that there is an effective mass term $\propto f''(G)G'$ appearing in the gravitational wave equation of motion.\\


In the context of spatially flat $f(G)$ cosmology, our findings reveal that in vacuum de Sitter space emerges as the sole admissible background for any choice of background connection. To depart from it we had to include matter, here in the form of a scalar field $\Xi$. Unfortunately, we observe that only the two gravitational tensor modes propagate, while the gravitational scalar and vector perturbations can be integrated out even in generic non-de Sitter cosmologies. Hence, we have only the two gravitational wave modes $h_{ij}$ propagating - with a new $1/\sqrt{f'}$ decay - in addition to the propagating scalar field $\delta\Xi$ of the matter field. The theory suffers from strong coupling problem for generic FLRW spacetimes. In the spatially curved case we find that we can still integrate out all perturbations except the two metric tensor modes, but for connections 1 and 3 we find a non-vanishing equation of motion for the axial scalar $b$, which fixes $b$ algebraically. For connection 3 this also happens for the connection tensor mode $h_{cij}$. These variables were undetermined in the spatially flat geometry, which separates these branches of solutions from each other and one cannot go from one branch to the other one in a smooth way.

For our choice of matter there is no hypermomentum, which for flat connections 2 and 3 and curved connection 2 lead precisely to constant $G$, so a direct coupling of matter to the teleparallel connection - and hence having non-vanishing hypermomentum - might lead to non-constant $G$ and more interesting dynamics of the perturbations. Similarly, the presence of hypermomentum would lead in general to the connection function $\Lie$ having a more interesting form, which might lead to more interesting solutions.

\section{Metric teleparallel theory}

Similarly to the general teleparallel case we now consider perturbations of a metric-teleparallel theory and derive the action of the perturbations. The action we consider is of the form
\begin{equation}\label{eq:MTAction}
    S=S_{MT}+S_\lambda+S_M=\int\dd^dx\left(\sqrt{-g}\Lie(g_{\mu\nu},T^\alpha{}_{\mu\nu})+\lambda_\alpha{}^{\beta\mu\nu}R^\alpha{}_{\beta\mu\nu}+\lambda^{\alpha\mu\nu}Q_{\alpha\mu\nu}\right)+S_M\,,
\end{equation}
where $S_\lambda$ contains both Lagrange multiplier parts. Let us again derive the background equations of motions briefly. Since the non-metricity tensor depends on both metric and connection the metric equation of motion becomes
\begin{equation}
    \E_{\mu\nu}+(\nabla_\alpha+T_\alpha)\lambda^\alpha{}_{\mu\nu}=T_{\mu\nu}\,,
\end{equation}
where now
\begin{equation}
    \E_{\mu\nu}=\frac{1}{\sqrt{-g}}\frac{\delta S_{MT}}{\delta g^{\mu\nu}}\,.
\end{equation}
Varying \eqref{eq:MTAction} w.r.t. the connection leads to
\begin{equation}\label{eq:fTConnEOM}
    \sqrt{-g}(Y_\alpha{}^{\mu\nu}-\HH_\alpha{}^{\mu\nu})-2(\nabla_\rho+T_\rho)\lambda_\alpha{}^{\nu\rho\mu}+T^\mu{}_{\rho\sigma}\lambda_\alpha{}^{\nu\rho\sigma}-2\lambda^{\mu\nu}{}_\alpha=0\,,
\end{equation}
with
\begin{equation}
    Y_\alpha{}^{\mu\nu}=\frac{1}{\sqrt{-g}}\frac{\delta S_{MT}}{\delta\Gamma^\alpha{}_{\mu\nu}}\,.
\end{equation}
Taking a divergence of \eqref{eq:fTConnEOM} one finds
\begin{equation}\label{eq:MTConnectionVariation1}
    (\nabla_\mu+T_\mu)\left(\sqrt{-g}(Y_\alpha{}^\mu{}_\nu-\HH_\alpha{}^\mu{}_\nu-2\lambda^\mu{}_{\nu\alpha}\right)=0\,.
\end{equation}
Since $\lambda_{\alpha\mu\nu}$ is symmetric in the last two indices we can use the symmetric part of \eqref{eq:MTConnectionVariation1} to replace the Lagrange multiplier in the metric equation of motion, leading to its final form
\begin{equation}
    \E_{\mu\nu}+\frac12(\nabla_\alpha+T_\alpha)\left(\sqrt{-g}(Y_{(\mu}{}^\alpha{}_{\nu)}-\HH_{(\mu}{}^\alpha{}_{\nu)})\right)=T_{\mu\nu}\,,
\end{equation}
and the connection equation of motion is simply the antisymmetric part of \eqref{eq:MTConnectionVariation1},
\begin{equation}
    \C_{\alpha\nu}=(\nabla_\mu+T_\mu)\left(\sqrt{-g}(Y_{[\alpha}{}^\mu{}_{\nu]}-\HH_{[\alpha}{}^\mu{}_{\nu]}\right)=0\,.
\end{equation}
For the Bianchi identity one can start from\footnote{The contribution from the Lagrange multipliers $S_\lambda$ again vanishes.} \eqref{eq:TPBianchi}, and then split the connection equation of motion into symmetric and antisymmetric parts. One can then show that this can be rewritten as
\begin{equation}
    2\sqrt{-g}\D^\mu\left(\E_{\mu\nu}+\frac12(\nabla_\alpha+T_\alpha)\left(\sqrt{-g}(Y_{(\mu}{}^\alpha{}_{\nu)}-\HH_{(\mu}{}^\alpha{}_{\nu)})\right)-T_{\mu\nu}\right)+(\nabla_\mu+T_\mu)C_\nu{}^\mu+T^\alpha{}_{\mu\nu}C_\alpha{}^\mu=0\,,
\end{equation}
which has schematically the same form as \eqref{eq:TPBianchi}.

\subsection{Perturbation action}
Similarly as above for the teleparallel action we derive the action of the perturbations by looking at the second order action. The Lagrange multiplier part $S_\lambda$ gives now a contribution, after manipulating the $\lambda_\alpha{}^{\beta\mu\nu}$ part similarly as when deriving \eqref{eq:TPPerturbationAction},
\begin{align}
    \delta^{(2)}S_\lambda&=-\int\dd^dx\,\left(\sqrt{-g}(Y_\alpha{}^{\nu\beta}-\HH_\alpha{}^{\nu\beta}-2\lambda^{\nu\beta}{}_\alpha\right)\nabla_\nu\delta\Lambda^\lambda{}_\beta\delta\Lambda^\alpha{}_\lambda+2\lambda_\alpha{}^{\mu\nu}\delta\Gamma^\lambda{}_{\alpha\mu}\delta g_{\nu\lambda}\,.
\end{align}
Using $\delta g_{\mu\nu}=2\delta\Lambda_{(\mu\nu)}$ the part containing $\lambda^{\alpha\mu\nu}$ becomes
\begin{align}
    &2\int\dd^dx\,\lambda^{\alpha\mu\nu}\nabla_\alpha\delta\Lambda^\lambda{}_\mu \left(\delta\Lambda_{\nu\lambda}-2\delta\Lambda{}_{(\nu\lambda)}\right)=-2\int\dd^dx\,\lambda^{\alpha\mu\nu}\nabla_\alpha\delta\Lambda^\lambda{}_\mu\delta\Lambda_{\lambda\nu}=\nonumber\\
    =&-\int\dd^dx\,\lambda^{\alpha\mu\nu}\nabla_\alpha(\delta\Lambda^\lambda{}_\mu\delta\Lambda_{\lambda\nu})=\int\dd^dx\,(\nabla_\alpha+T_\alpha)\lambda^{\alpha\mu\nu}\delta\Lambda^\lambda{}_\mu\delta\Lambda_{\lambda\nu}=\nonumber\\
    =&\int\dd^dx\,(\nabla_\alpha+T_\alpha)\left(\sqrt{-g}(Y^{(\nu|\alpha|\mu)}-\HH^{(\nu|\alpha|\mu)})\right)\delta\Lambda^\lambda{}_\mu\delta\Lambda_{\lambda\nu}\,,
\end{align}
so the total action of the perturbations is
\begin{align}
    \delta^{(2)}S&=\delta^{(2)}S_{MT}-\int\dd^dx\,\Big[\sqrt{-g}(Y_\alpha{}^{\nu\beta}-\HH_\alpha{}^{\nu\beta})\nabla_\nu\delta\Lambda^\lambda{}_\beta\delta\Lambda^\alpha{}_\lambda-\nonumber\\
    &-\left.(\nabla_\alpha+T_\alpha)\left(\sqrt{-g}(Y^{(\nu|\alpha|\mu)}-\HH^{(\nu|\alpha|\mu)})\right)\delta\Lambda^\lambda{}_\mu\delta\Lambda_{\lambda\nu}\right]+\delta^{(2)}S_M\,.
\end{align}
$S_M$ and its perturbations are as for $f(G)$ cosmology. The 16 perturbation variables are \\$(\phi,j,\psi,\sigma,S^i,F^i,h_{ij},e,E^i,b,B^i)$, in addition to $\delta\Xi$ for the matter field.

\subsection{Flat $f(\T)$ cosmology}
Another simple extension of the GR equivalent teleparallel theory is to use the Lagrangian $\Lie=f(\T)$, with $f$ again a smooth function. There is only one admissible connection compatible with cosmological symmetries and a spatially flat background \cite{Hohmann:2017jao,DAmbrosio2021:2109.04209v2}, with only $\Gamma^\mu{}_{0\nu}=\HH\delta^\mu_\nu$ non-zero. This connection has no axial component, $\epsilon^{ijk}\Gamma_{ijk}=0$. We then have $\T=6\HH^2/a^2$, and the background equations of motion lead to the same form as in $f(G)$ cosmology for connection 1, with $\T$ instead of $G$. The background is thus again de Sitter in the case of the absence of matter.  

In the scalar sector the axial perturbation $b$ is absent and $\E_b=0$. We can solve $\E_\phi=0$ for $\phi$, $\E_j=0$ for $j$, and $\E_e=0$ for $e$, followed by the replacement
\begin{equation}
\delta\Xi\to\delta\Xi+\frac{\Xi'}{3\HH}(k^2\sigma-3\psi)\,.
\end{equation}
We then have the solutions
\begin{dmath}
    {\phi=\ }\frac{1}{12 \mathcal{H}^2 f' \left(a^2 f'+12 \mathcal{H}^2 f''\right)}\Big(a^2 f' \left(\Xi '\right)^2 \left(k^2 \sigma -3 \psi \right)+a^2 \mathcal{H} f' \left(3 \delta \Xi  \Xi '+4 f' \left(k^2 \sigma '-3 \psi '\right)\right)+12 \mathcal{H}^3 f'' \left(3 \delta \Xi  \Xi '+4 f' \left(k^2 \sigma '-3 \psi '\right)\right)\,,\\
    {j=\ }\frac{\frac{a^2 \delta \Xi  \mathcal{V}'(\Xi )+\delta \Xi ' \Xi '+4 k^2 \psi  f'}{4 k^2 f'}-\frac{k^2 \sigma }{3}}{\mathcal{H}}+\frac{3 \delta \Xi  \Xi '}{4 k^2 f'}-\frac{\delta \Xi  \left(\Xi '\right)^3}{16 k^2 \mathcal{H}^2 \left(f'\right)^2}+\sigma '\,,\\
    {e=\ }\frac{1}{24 k^2 \mathcal{H}^2 \left(f'\right)^2}\Big(4 \mathcal{H} f' \left(3 \left(a^2 \delta \Xi  \mathcal{V}'(\Xi )+\delta \Xi ' \Xi '+4 k^2 \psi  f'\right)-4 k^4 \sigma  f'\right)-3 \delta \Xi  \left(\Xi '\right)^3\Big)\,,
\end{dmath}
and $\E_\psi=\frac{\Xi'}{\HH}\E_{\delta\Xi}$ and $\E_\sigma=-\frac{k^2\Xi'}{3\HH}\E_{\delta\Xi}$. We have the same equation of motion for $\delta\Xi$ as for $f(G)$ cosmology with connection 1, as well as the solutions for the Bardeen potentials $\Phi$ and $\Psi$. In fact the vector and tensor sectors in $f(\T)$ are also just as in $f(G)$ cosmology with connection 1, except that of course we only have the $E^i,B^i$ connection perturbations. Still, the equations of motion and solutions for the $f(\T)$ variables are just like in $f(G)$ cosmology with connection 1, so the same conclusions drawn above apply, and we have again only $\delta\Xi$ as a scalar degree of freedom, and no gravitational ones beside the two tensor modes. Hence, $f(\T)$ theories suffer from the same strong coupling problem for generic cosmological backgrounds, yielding them problematic for cosmological applications. 


\subsection{Curved $f(\T)$ cosmology}

In the spatially curved case we have two admisseble background connections\footnote{Up to the replacement $u\to-u$ similar to the curved $f(G)$ connections.}, namely connection 1 with $\Gamma^\mu{}_{0\nu}=\HH\delta^\mu_\nu$ and $\Gamma^i{}_{jk}$ as for connection 1 of $f(G)$ cosmology with $\K=0$. The background equations of motion are now as for connection 1 in $f(G)$ gravity, with $\T=G=6(\HH^2-u^2)/a^2$; note that this connection has an axial component. Connection 2 has the same form as connection 3 from curved $f(G)$ gravity, with again $\K=0$. Its background equations also have the same form.

\subsubsection{Connection 1}

In the scalar sector we find $\E_b$ now not vanishing; in fact it has the same form as in $f(G)$ gravity with $j_c=j/2$ - which is demanded in $f(\T)$ gravity by metric-compatibility - leading to the same algebraic solution for $b$. We can then solve $\E_{\phi,j,e}=0$ for $\phi,j,e$ as in the flat case, and after similar manipulations as for $f(G)$ cosmology we find only $\delta\Xi$ propagating. The final solutions for $\E_{\delta\Xi},\phi,j,e,\Phi,\Psi$ and $b$ also have the same form as in $f(G)$ cosmology with connection 1 with $G=T=6\HH^2/a^2$, including the divergence in the flat limit, $b\propto 1/u$, which again signals that these two branches of solutions are disconnected. 

In the vector and tensor sector we also find the same solutions as in $f(G)$ cosmology with connection 1, with the perturbations $S^i_c,F^i_c,h_{cij}$ of course here already fixed to their metric counterparts over two.

\subsubsection{Connection 2}

In the scalar sector we solve $\E_{\phi,j,e}=0$ for $\phi,j,e$. We then find that the results for $\E_{\delta\Xi},\phi,j,e,\Phi,\Psi,$ agree with those of connection 3 of $f(G)$ cosmology with $\T=G$. In particular $\E_b$ has the same form. In the vector and tensor sector we again have the same solutions as in $f(G)$ cosmology with connection 3, but of course $h_{cij}=h_{ij}/2$ is fixed already in $f(\T)$ cosmology.\\

We conclude that in $f(\T)$ cosmology the background and perturbation equations have - after solving them - the same solutions as in $f(G)$ cosmology, with the appropriate identification of connections. This is not surprising, since while the only difference to $f(G)$ cosmology lies in the $f(\T)$ background connections being more fixed, with no free functions remaining, in the special case of $f(G)$ cosmology the free connection function $\Lie$ was fixed by the connection equations of motion, while the function $\K$ never appears at the background level and is unfixed. $f(G)$ cosmology thus offers a wider range of possible connections only up to the undetermined function $\K$, which does not affect the dynamics of either background or perturbations; $f(\T)$ could thus reproduce $f(G)$ results to first order. We also found the additional connection perturbations $\delta\Lambda_{(\mu\nu)}$ that are available in $f(G)$ gravity not contributing to the solutions of the perturbations, so no differences between $f(G)$ and $f(\T)$ gravity appears in the final solutions of the perturbations. Unfortunately, all the generic cosmological solutions in $f(\T)$ theories suffer from strong coupling problems in the same way as in $f(G)$ theories and this result persists even after the inclusion of matter fields and curved spacetimes.

\section{Symmetric teleparallel theory}
Finally we turn to symmetric teleparallism, and derive again the action of the pertubations $\delta^{(2)}S$. The action now has the general form
\begin{equation}\label{eq:STAction}
    S=S_{ST}+S_\lambda+S_M=\int\dd^dx\left(\sqrt{-g}\Lie(g_{\mu\nu},Q_{\alpha\mu\nu})+\lambda_\alpha{}^{\beta\mu\nu}R^\alpha{}_{\beta\mu\nu}+\lambda_\alpha{}^{\mu\nu}T^\alpha{}_{\mu\nu}\right)+S_M\,,
\end{equation}
with $S_\lambda$ again the action containing the Lagrange multipliers. Since Riemann and torsion tensors depend only on the connection the metric equations of motion are schematically of the form
\begin{equation}
    \E_{\mu\nu}=T_{\mu\nu}\,,
\end{equation}
with
\begin{equation}
    \E_{\mu\nu}=\frac{1}{\sqrt{-g}}\frac{\delta S_{ST}}{\delta g^{\mu\nu}}\,.
\end{equation}
Varying \eqref{eq:STAction} w.r.t. to the connection leads to
\begin{equation}
    \sqrt{-g}(Y_\alpha{}^{\mu\nu}-\HH_\alpha{}^{\mu\nu})-2\nabla_\rho\lambda_\alpha{}^{\nu\rho\mu}+2\lambda_\alpha{}^{\mu\nu}=0\,,
\end{equation}
with
\begin{equation}
    Y_\alpha{}^{\mu\nu}=\frac{1}{\sqrt{-g}}\frac{\delta S_{ST}}{\delta\Gamma^\alpha{}_{\mu\nu}}\,.
\end{equation}
To eliminate both Lagrange multipliers we take divergences $\nabla_\mu$ and $\nabla_\nu$, leading to
\begin{equation}
    \C_\alpha=\nabla_\mu\nabla_\nu\left(\sqrt{-g}(Y_\alpha{}^{\mu\nu}-\HH_\alpha{}^{\mu\nu})\right)=0\,,
\end{equation}
which is now a third order differential equation. The Bianchi identity can be straightforward derived using now
\begin{equation}
    \Lie_\xi\Gamma^\alpha{}_{\mu\nu}=\nabla_\mu\nabla_\nu\xi^\alpha
\end{equation}
to be
\begin{equation}
    2\sqrt{-g}\D^\mu(\E_{\mu\nu}-T_{\mu\nu})+\C_\nu=0\,,
\end{equation}
so we see that the connection equation of motion is proportional to the divergence of the metric equation of motion. Once the metric equations of motion are fulfilled, $\E_{\mu\nu}=T_{\mu\nu}$, the connection ones will be automatically as well, $\C_\mu=0$. Hence they may be discarded, and only metric ones - which are manifestly second order differential equations - are used.

\subsection{Perturbation action}
Lastly we need to derive the action for the perturbations. As in \eqref{eq:TPPerturbationAction} we have now the contribution
\begin{align}
    -&\int\dd^dx\,\sqrt{-g}(Y_\alpha{}^{\nu\beta}-\HH_\alpha{}^{\nu\beta})\nabla_\nu\delta\Lambda^\lambda{}_\beta\delta\Lambda^\alpha{}_\lambda=\nonumber\\
    =&-\int\dd^dx\,\sqrt{-g}(Y_\alpha{}^{\nu\beta}-\HH_\alpha{}^{\nu\beta})\nabla_\nu\nabla_\beta\delta\rchi^\lambda\nabla_\lambda\delta\rchi^\alpha\,,
\end{align}
from the Lagrange multiplier part $S_\lambda$, leading to the full perturbation action
\begin{equation}
    \delta^{(2)}S=\delta^{(2)}S_{ST}-\int\dd^dx\,\sqrt{-g}(Y_\alpha{}^{\nu\beta}-\HH_\alpha{}^{\nu\beta})\nabla_\nu\nabla_\beta\delta\rchi^\lambda\nabla_\lambda\delta\rchi^\alpha+\delta^{(2)}S_M\,.
\end{equation}
This action might potentially lead to third and fourth order time derivative terms of the connection perturbations variables through terms of the form
\begin{equation}
    \delta\Gamma^2\sim (\nabla^2\delta\rchi)^2\,,
\end{equation}
but recall that one can always make the perturbation equations of motion manifestly second order by choosing the perturbative coincident gauge with $\delta\rchi^\mu=0$. For example, in the cosmological setup we could pick the gauge with $\xi^0=-\Chi$, leading to $\Chi\to 0$ but also to
\begin{equation}
    \phi\to\phi-\Chi'-\HH\Chi\,.
\end{equation}
We thus see that higher time derivatives of $\Chi$ would automatically enter the perturbation equations of motion if $\phi$ enters with second derivatives. We can then try to use this to eliminate higher derivatives of $\Chi$ by replacing $\phi\to\phi+\Chi'$ in the perturbation action $\delta^{(2)}S$. It turns out that in our example this is sufficient to eliminate higher time derivatives of the connection perturbations. We call this the perturbative coincidence gauge trick. However, the perturbation equations of motion simplify significantly when using the full perturbative coincidence gauge trick for all variables in a similar way, i.e. we replace in the action $\delta^{(2)}S$
\begin{align}
    \phi&\to\phi+\Chi'+\HH\Chi\,,\\
    j&\to j+\Theta'-\Chi\,,\\
    \psi&\to\psi-\HH\Chi+\frac{k^2}{3}\Theta\,,\\
    \sigma&\to\sigma+\Theta\,,\\
    \delta\Xi&\to\delta\Xi+\Chi\Xi'
\end{align}
in the scalar sector, and
\begin{align}
    S^i&\to S^i+{\Upsilon^i}'\,,\\
    F^i&\to F^i+\Upsilon^i
\end{align}
in the vector sector. In our examples this renders all the perturbation equations of motion independent of $\Chi,\Theta,\Upsilon^i$, and also makes $\E_\Chi$, $\E_\Theta$, and $\E_{\Upsilon^i}$ vanish identically. Note in particular that since this transformation has a form of a gauge transformation for the metric variables alone, the gauge invariant Bardeen potentials $\Phi,\Psi$, and $V^i$ remain unaffected. Note that by this trick we do not fix a gauge, since we make the above replacement without setting $\delta\rchi^\mu$ to zero.

\subsection{Flat $f(\Q)$ cosmology}
We pick now the Lagrangian $\Lie=f(\Q)$. There are three possible connections compatible with spatially flat cosmological symmetries \cite{DAmbrosio2021:2109.04209v2}. Connection 1 has only $\Gamma^0{}_{00}=C\neq 0$, where $C=C(\eta)$ is an arbitrary function of time. Connection 2 has only $\Gamma^0{}_{ij}=C\delta_{ij}$ and $\Gamma^0{}_{00}=-C'/C$ non-vanishing. Connection 3 has $\Gamma^i{}_{0j}=C\delta^i_j$ and $\Gamma^0{}_{00}=C+C'/C$ non-vanishing.  The 14 perturbation variables are  $(\phi,j,\psi,\sigma,S^i,F^i,h_{ij},\Chi,\Theta,\Upsilon^i)$, in addition to the scalar field perturbation $\delta\Xi$.

Connection 1 has at the background level $\Q=6\HH^2/a^2$ and $C$ does not appear in the equations of motion. They take again the same form as in $f(G)$ cosmology with connection 1, with $\Q=G$. For connection 2 we have instead the more complicated vacuum equations
\begin{align}
    \Q&=\frac{6\HH^2-6\HH C-3C'}{a^2}\,,\\
    f(\Q)&=\frac{3f'(\Q)}{f'(\Q)a^4+3f''(\Q)a^2(C^2-4\HH^2)}\Big[f'(\Q)a^2(4\HH^2-2\HH C-C')-\nonumber\\
    &-3f''(\Q)\left(16\HH^4-8\HH^3 C+2\HH C^3-2\HH^2(3C^2+2C')+C(CC'+C'')\right)\Big]\,,\\
    \HH'&=\frac{2f'(\Q)a^2\HH^2-3f''(\Q)(2\HH+C)(4\HH^2(\HH-C)+C'')}{2f'(\Q)a^2+6f''(\Q)(C^2-4\HH^2)}\,.
\end{align}
This shows that even in vacuum the connection may result in non-trivial background dynamics, and are in general not just de Sitter. Connection 3 has similar background equations,
\begin{align}
    \Q&=\frac{6\HH^2-6\HH C-3C'}{a^2}\,,\\
    f(\Q)&=\frac{3f'(\Q)}{f'(\Q)a^4+3f''(\Q)a^2(4\HH^2-8\HH C+3C^2)}\Big[f'(\Q)a^2(4\HH^2-2\HH C-C')+\nonumber\\
    &+3f''(\Q)(16\HH^4-40\HH^3 C+\HH^2(30C^2-4C')+\HH(-6C^3+8CC')-C(3CC'+C'')\Big]\,,\\
    \HH'&=\frac{2f'(\Q)a^2\HH^2+3f''(\Q)(2\HH-3C)(4\HH^2(\HH-C)+C'')}{2f'(\Q)a^2+6f''(\Q)(4\HH^2-8\HH C+3C^2)}\,.
\end{align}
Since the background dynamics are not fixed to de Sitter even in vacuum, i.e. $\HH\neq\HH^2$ in general, we consider\footnote{We also checked the results below when including the scalar field $\Xi\neq0$, but it only made the equations more complicated and did not lead to qualitatively different results.} only vacuum $S_M=0$ for connections 2 and 3, i.e. we set $\Xi=\delta\Xi=0$. For connections 2 and 3 we can however recover de Sitter dynamics if we choose
\begin{equation}\label{eq:deSitterC}
    C=\frac{C_0}{a^2}\,,
\end{equation}
with $C_0$ a constant. This leads to $\Q=6\HH^2/a^2$, and $\HH'=\HH^2$. Note that taking the limit $C_0\to 0$ carefully, connections 2 and 3 return to the form of connection 1, so we consider $C_0\neq 0$ only. In this case we apply the background equations of motion by solving them for $f(\Q)$ and $\HH'$, leading to de Sitter like dynamics with
\begin{align}
    f&=\frac{12f'\HH^2}{a^2}\,,\\
    \HH'&=\HH^2\,,
\end{align}
with $f=f(\Q)$ and similar for $f',f''$. In the case of general $C$ it turns out that for connections 2 and 3 the perturbations equations take a simpler form if we instead solve for $C''$ and $\HH'$, e.g. for connection 2 we set
\begin{dmath}
    {C''=\ }\frac{1}{9 C}\Big(-\frac{3 a^2 f \left(C^2-4 \mathcal{H}^2\right)}{f'}+\frac{a^2 \left(a^2 f+3 f' \left(C'+2 \mathcal{H} (C-2 \mathcal{H})\right)\right)}{f''}-9 \left(\left(C^2-4 \mathcal{H}^2\right) C'+2 \mathcal{H} \left(-3 C^2 \mathcal{H}+C^3-4 C \mathcal{H}^2+8 \mathcal{H}^3\right)\right)\Big)\,,\\
    {\HH'=\ }\frac{1}{6 C}\Big(\frac{a^2 f (C+2 \mathcal{H})}{f'}+6 \mathcal{H} \left(C'+C^2\right)+3 C C'+6 C \mathcal{H}^2-24 \mathcal{H}^3\Big)\,,
\end{dmath}
and for connection 3
\begin{dmath}
    {C''=\ }\frac{1}{9 C}\Big(-\frac{3 a^2 f \left(3 C^2-8 C \mathcal{H}+4 \mathcal{H}^2\right)}{f'}-\frac{3 a^2 f' \left(C'+2 \mathcal{H} (C-2 \mathcal{H})\right)+a^4 f}{f''}+18 \mathcal{H} (\mathcal{H}-C) \left(3 C^2-12 C \mathcal{H}+8 \mathcal{H}^2\right)-9 (2 \mathcal{H}-3 C) (2 \mathcal{H}-C) C'\Big)\,,\\
    {\HH'=\ }\frac{1}{6 C}\Big(\frac{a^2 f (3 C-2 \mathcal{H})}{f'}+6 \mathcal{H} \left(3 C^2-C'\right)+9 C C'-42 C \mathcal{H}^2+24 \mathcal{H}^3\Big)\,.
\end{dmath}

\subsubsection{Connection 1}

In the scalar sector we solve $\E_\phi=0$ for $\phi$ and $\E_j=0$ for $j$,
which give very long expressions that we do not report here. One then finds that the kinetic matrix\footnote{If we enumerate the remaining perturbation variables $v^i$ then the kinetic matrix $K_{ij}$ contains the coefficients of ${v^i}''$ in $\E_{v^j}$. The kinetic part of the Lagrangian is then simply $-\frac12 K_{ij}{v^i}'{v^j}'$. If the kinetic matrix has full rank then all variables $v^i$ are dynamical in the sense that we can solve the equations of motion for all ${v^i}''$.} $K_{ij}$ for the three remaining variables $\psi,\sigma,\delta\Xi$ has the form
\begin{equation}
    K_{ij}=\frac1b\begin{pmatrix}
        c & d & e \\ d & h & l \\ e & l & m
    \end{pmatrix}\,,
\end{equation}
with
\begin{dmath}
    {b=\ }\Big(a^6 \left(f'\right)^2 f'' \left(12 \mathcal{H}^2 \left(-C^2-4 C \mathcal{H}+44 \mathcal{H}^2\right) f'+\left(C^2-4 C \mathcal{H}-20 \mathcal{H}^2\right) \left(\Xi '\right)^2\right)+3 a^4 \mathcal{H}^2 \left(f''\right)^2 \left(8 \left(C^2-C \mathcal{H}-14 \mathcal{H}^2\right) f' \left(\Xi '\right)^2+96 \mathcal{H}^2 \left(-C^2-4 C \mathcal{H}+20 \mathcal{H}^2\right) \left(f'\right)^2+3 \left(\Xi '\right)^4\right)+144 a^2 \mathcal{H}^4 (2 \mathcal{H}-C) \left(f''\right)^3 \left(12 \mathcal{H}^2 (C+6 \mathcal{H}) f'-(C+4 \mathcal{H}) \left(\Xi '\right)^2\right)+16 a^8 \mathcal{H}^2 \left(f'\right)^4\Big)\,,\\
    {c=\ }\Big(2 \left(\Xi '\right)^2 \left(-9 a^2 \mathcal{H}^2 f' \left(f''\right)^2 \left(8 \left(C^2+4 C \mathcal{H}-20 \mathcal{H}^2\right) f'+3 \left(\Xi '\right)^2\right)+3 a^4 \left(-C^2-4 C \mathcal{H}+44 \mathcal{H}^2\right) \left(f'\right)^3 f''+4 a^6 \left(f'\right)^4-108 \mathcal{H}^4 \left(f''\right)^3 \left(3 \left(\Xi '\right)^2-4 (2 \mathcal{H}-C) (C+6 \mathcal{H}) f'\right)\right)\Big)\,,\\
    {d=\ }\frac{1}{3}\Big(-8 k^2 f' \left(\Xi '\right)^2 \left(a^2 f'+12 \mathcal{H}^2 f''\right)^2 \left(a^2 f'-3 \mathcal{H} (C+\mathcal{H}) f''\right)\Big)\,,\\
    {e=\ }\Big(2 \mathcal{H} \Xi ' \left(3 a^4 \left(f'\right)^2 f'' \left(\left(-C^2-4 C \mathcal{H}+44 \mathcal{H}^2\right) f'-\left(\Xi '\right)^2\right)+9 a^2 \mathcal{H} f' \left(f''\right)^2 \left(8 \mathcal{H} \left(-C^2-4 C \mathcal{H}+20 \mathcal{H}^2\right) f'+(C-6 \mathcal{H}) \left(\Xi '\right)^2\right)+4 a^6 \left(f'\right)^4+108 \mathcal{H}^3 (2 \mathcal{H}-C) \left(f''\right)^3 \left(4 \mathcal{H} (C+6 \mathcal{H}) f'-\left(\Xi '\right)^2\right)\right)\Big)\,,\\
    {h=\ }\frac{1}{9}\Big(2 k^4 f' \left(-3 a^4 \left(f'\right)^2 f'' \left(12 \mathcal{H}^2 (C+2 \mathcal{H})^2 f'-\left(C^2-4 C \mathcal{H}+12 \mathcal{H}^2\right) \left(\Xi '\right)^2\right)-9 a^2 \mathcal{H}^2 \left(f''\right)^2 \left(96 \mathcal{H}^2 (C+2 \mathcal{H})^2 \left(f'\right)^2+8 (C+2 \mathcal{H}) (3 \mathcal{H}-C) f' \left(\Xi '\right)^2-3 \left(\Xi '\right)^4\right)+4 a^6 \left(f'\right)^3 \left(\Xi '\right)^2-432 \mathcal{H}^4 \left(f''\right)^3 \left(12 \mathcal{H}^2 (C+2 \mathcal{H})^2 f'+(2 \mathcal{H}-C) (C+4 \mathcal{H}) \left(\Xi '\right)^2\right)\right)\,,\\
    {l=\ }\frac{1}{3}\Big(-2 k^2 \mathcal{H} f' \Xi ' \left(a^2 f'+12 \mathcal{H}^2 f''\right) \left(3 a^2 f'' \left(4 \mathcal{H} (6 \mathcal{H}-C) f'-\left(\Xi '\right)^2\right)+4 a^4 \left(f'\right)^2+144 \mathcal{H}^3 (2 \mathcal{H}-C) \left(f''\right)^2\right)\Big)\,,\\
    {m=\ }\frac{1}{2}\Big(\mathcal{H}^2 \left(12 a^4 \left(f'\right)^2 f'' \left(\left(-C^2-4 C \mathcal{H}+44 \mathcal{H}^2\right) f'-2 \left(\Xi '\right)^2\right)+9 a^2 \left(f''\right)^2 \left(32 \mathcal{H}^2 \left(-C^2-4 C \mathcal{H}+20 \mathcal{H}^2\right) \left(f'\right)^2+8 \mathcal{H} (C-6 \mathcal{H}) f' \left(\Xi '\right)^2+\left(\Xi '\right)^4\right)+16 a^6 \left(f'\right)^4+864 \mathcal{H}^3 (2 \mathcal{H}-C) \left(f''\right)^3 \left(2 \mathcal{H} (C+6 \mathcal{H}) f'-\left(\Xi '\right)^2\right)\right)\Big)\,.
\end{dmath}

$K_{ij}$ is found to have the property
\begin{equation}\label{eq:fQConn1ScalarK}
    \det(K_{ij})\propto (\HH'-\HH^2)^2 (f'')^2\,,
\end{equation}
where ${\Xi'}^2\propto (\HH'-\HH^2)$ was used. We find that $K_{ij}$ has full rank as long as we are not in de Sitter space, $\HH'=\HH^2$, or STEGR, $f''=0$. In case of $\det(K_{ij})\neq0$ we are left with two propagating gravitational perturbations $(\psi,\sigma)$ in addition to the propagating scalar field $\delta\Xi$. We remark that for $C\to0$ the above $K_{ij}$ recovers the result from \cite{BeltranJimenez:2019tme}, up to a $(9-5\epsilon)$ instead of a $(2-5\epsilon)$ in $g$ in formula $(\text{B}1)$.

The eigenvalues of $K_{ij}$ are unfortunately very lengthy expressions that do not provide much insight. Whether all of them are positive - a requirement to avoid ghost degrees of freedom - is not clear. Similarly, it is not obvious that $\det(K_{ij})>0$, a necessary requirement for the absence of ghosts. This has to be checked for concrete solutions of the background equations of motion. However, the equations of motion of the remaining perturbations, as can be seen by the kinetic matrix above, still depend on the connection $C$ that could not be determined by the background equations of motion. Even the equations of motion for the gauge invariant Bardeen potentials depend on $C$, which in turn makes their solutions arbitrary since $C$ is. This inconsistency may be attributed to the choice of connection and high symmetry of the background, which makes $C$ disappear from the background equations of motion.

\eqref{eq:fQConn1ScalarK} shows however that in de Sitter the kinetic matrix has no longer full rank, which we can achieve by setting $\Xi=0$ at the background level. Hence, the spatially flat de Sitter spacetimes suffer from strong coupling problem. Fixing the background to de Sitter in this way we can additionally solve $\E_\psi=0$ for $\psi$, leading to $\E_\sigma=0$ identically, and the solutions
\begin{dmath}
    {\phi=\ }\frac{1}{(C-2 \mathcal{H})^2}\Big(\mathcal{H} \sigma ' \left(-4 C'+C^2+4 C \mathcal{H}-4 \mathcal{H}^2\right)+\left(C^2-4 \mathcal{H}^2\right) \sigma ''\Big)\,,\\
    {j=\ }\frac{4\HH\sigma'}{2\HH-C}\,,\\
    {\psi=\ }\frac{k^2}{3}\sigma+\HH\frac{2\HH+C}{2\HH-C}\sigma'\,.\\
\end{dmath}
This leads to $\Phi=\Psi=0$, so the gravitational scalar perturbations vanish and decouple from $\delta\Xi$. We see that here, while the metric perturbations depend on the undetermined connection function $C$, the physical gauge invariant potentials $\Phi$ and $\Psi$ do not. Hence the $C$ dependence of the metric perturbation variables may be gauged away. All that is left now is the standard cosmological equation of motion for $\delta\Xi$,
\begin{equation}
    \E_{\delta\Xi}=\frac{1}{2a^2}\left(\delta\Xi''+2\HH\delta\Xi'+(k^2+m^2a^2)\delta\Xi\right)=0\,,
\end{equation}
where we defined $m^2=\mathcal{V}''(0)$ as the effective mass of $\delta\Xi$. We have integrated out all gravitational scalar perturbations, and are left with no propagating perturbations except $\delta\Xi$.
The vector and tensor sectors are as for $f(G)$ cosmology with connection 1, so we have two propagating tensor modes, and the vector modes fixed with $V^i=0$. This holds for both de Sitter and general backgrounds.

We find that, unlike $f(G)$ and $f(\T)$ cosmology, we do have five propagating degrees of freedom in non-de Sitter backgrounds for connection 1, namely two scalars $\psi,\sigma$ in addition to the two tensor modes and $\delta\Xi$. However, the other three modes in the gravity sector become strongly coupled. Going to de Sitter space by setting $\Xi=0$ allows us to further integrate out $\psi$ and $\sigma$, suggesting an even worse strong coupling problem of de Sitter space for this background connection in the scalar sector.

\subsubsection{Connection 2}

Let us start with general $C$ first. Solving $\E_j=0$ algebraically for $j$ leads to
\begin{dmath}
    j=\frac{1}{6 a^2 C^2 f' \left(C'+2 C \mathcal{H}-4 \mathcal{H}^2\right)+2 a^4 C^2 f+3 k^2 f'' \left(C'+2 C \mathcal{H}\right)^2}\Big(C \left(a^2 f' \left(-9 \psi  C'-3 \phi  C'-4 C k^2 \sigma '+12 C \psi '+6 C \mathcal{H} (\phi -3 \psi )+12 \mathcal{H}^2 (3 \psi +\phi )\right)+a^4 (-f) (3 \psi +\phi )+9 f'' \left(C'+2 C \mathcal{H}\right) \left(2 \psi  C'+4 \mathcal{H} \left(C \psi +\psi '\right)+C \left(\phi '-\psi '\right)+4 \mathcal{H}^2 \phi \right)\right)\Big)\,.
\end{dmath}
The three remaining equations $\E_\phi$, $\E_\psi$, and $\E_\sigma$ have a kinetic matrix of rank three,
\begin{align}
    K_{ij}&=\frac{1}{b}\begin{pmatrix} c & d & e \\ d & -3h & l \\  e & l & m \end{pmatrix}\,,\\
    b&=2fC^2a^6+6f'C^2a^4(C'+2\HH C-4\HH^2)+3f''k^2a^2(C'+2\HH C)^2\,,\\
    c&=-9C^4 f''(fa^2+3f'(C'+2\HH C-4\HH^2)\,,\\
    d&=9f''C^2\left(fCa^2(C-4\HH)+2f'C\HH(2k^2+3(C^2-6\HH C+8\HH^2))+f'C'(2k^2+3C(C-4\HH))\right)\,,\\
    e&=-6k^4C^2 f'f''(C'+2\HH C)\,,\\
    h&=4ff''C^2a^4+3f'f''\Big(2\HH C^2(-96\HH^3+4k^2 C+96\HH C+3C^3-6\HH(2k^2+5C^2))+\\
    &+C(4k^2C+48\HH^2C+3C^3-8\HH(k^2+3C^2))C'+2k^2{C'}^2\Big)+\\
    &+C^2a^2\left(3ff''(C-4\HH)^2+4{f'}^2(3C'+6\HH C-12\HH^2-2k^2)\right)\,,\\
    l&=-2k^4Cf'(4f'Ca^2+3f''(4\HH-C))C'+2\HH C))\,,\\
    m&=\frac23f'k^4\left(2fC^2a^4+2f'C^2a^2(2k^2+3C'+6\HH C-12\HH^2)+3f''k^2(C'+2\HH C)^2\right)\,.
\end{align}
The determinant is given by
\begin{align}
    \det(K_{ij})=\frac{36(f')^2 f'' k^4C^4A}{a^4b}\,,
\end{align}
with
\begin{equation}
    A=fa^2+3f'(2\HH(-2\HH+C)+C')\,.
\end{equation}
This determinant is in general non-zero, but vanishes correctly for $f''=0$. We thus have three scalar degrees of freedom.

In the vector sector we solve $\E_{S^i}$ algebraically for $S^i$,
\begin{dmath}
    S^i=\frac{1}{3 f' \left(-2 C'-4 C \mathcal{H}+k^2+8 \mathcal{H}^2\right)-2 a^2 f}\Big(3 k^2 f' {F^i}'\Big)\,.
\end{dmath}
We then have a second order differential equation for $F^i$,
\begin{equation}
    \frac{f'k^2A}{a^2(2A-3f'k^2)}\left(F_i''-F_i'\frac{2A^2-12f'A\HH C-9(f')^2k^2C'}{3f'C(2A-3f'k^2)}-F_i\frac{2A-3f'k^2}{3f'}\right)=0\,,
\end{equation}
Note that this equation is only non-trivial for $A\neq 0$, otherwise $\E_{F^i}=0$ trivially. In fact for both scalar and vector sector to be fully dynamical we need $f''A\neq 0$. We thus have two vector degrees of freedom $F^i$.
In the tensor sector the equations for $h_{ij}$ read
\begin{equation}\label{eq:fQConn2TensorEquations}
    \frac{f'}{2a^2}\left(h_{ij}''+(2\HH+\ln(f'(\Q))')h_{ij}'+(k^2+2C\ln(f'(\Q))'h_{ij}\right)=0
\end{equation}
so while they do not take the same form as in GR we have still two tensor degrees of freedom. We now have an extra mass term $\propto Cf''\Q'$ that may affect the long wavelength modes $k^2\to0$, since $h_{ij}=$ const is no longer a solution. 

Let us now consider the case $C=C_0/a^2$ with $C_0\neq 0$, leading to the de Sitter dynamics governed by $\HH'=\HH^2$. In the scalar sector the above kinetic matrix $K_{ij}$ has no longer full rank, as one can see that $A=0$  and $\det(K_{ij})$ vanishes, but we begin again from the start with all equations of motion. The combination
\begin{equation}
    \frac{k^2(C_0-4\HH a^2)}{C_0}\E_\phi+\frac{k^2}{3}\E_j-\E_\psi'+\E_\sigma=0
\end{equation}
may then be solved for $j$,
\begin{dmath}
    {j=\ }\frac{1}{3 a^4 k^2 \mathcal{H} f' \left(4 a^2 \mathcal{H}-C_0\right)}\Big(a^4 \left(C_0 k^4 \sigma  f'-3 C_0 k^2 \psi  f'+3 C_0 \mathcal{H} f' \left(9 \psi '-4 k^2 \sigma '\right)+27 C_0 \mathcal{H}^2 \phi  f'+432 \mathcal{H}^4 f'' \psi '+432 \mathcal{H}^5 \phi  f''\right)+108 a^2 C_0 \mathcal{H}^3 f'' \left(\psi '+2 \mathcal{H} \phi +\phi '\right)+4 a^6 \mathcal{H} f' \left(k^4 (-\sigma )+3 k^2 \psi +9 \mathcal{H}^2 \phi +9 \mathcal{H} \psi '\right)+54 C_0^2 \mathcal{H}^2 f'' \left(\phi '-\psi '\right)\Big)\,.
\end{dmath}
Next the combination
\begin{equation}
    \frac{k^2}{12C_0}\left((C_0-16\HH a^2)\E_\phi+(C_0-4\HH a^2)\E_j\right)+\E_\psi=0
\end{equation}
can be solved algebraically for $\sigma'$, since $\sigma$ appears only with derivatives,
\begin{dmath}[spread={5pt}]
    {\sigma'=\ }\frac{1}{a^4 k^2 f' \left(-4 \mathcal{H}^2 \left(9 C_0^2-4 a^4 k^2\right)-8 a^2 C_0 k^2 \mathcal{H}-144 a^2 C_0 \mathcal{H}^3+C_0^2 k^2\right)}\Big(3 \left(a^4 C_0^2 k^2 f' \psi '+2 \mathcal{H}^3 \left(-72 a^6 C_0 f' \psi '-18 a^4 C_0^2 \phi  f'+8 a^8 k^2 \phi  f'+27 C_0^3 f'' \left(\psi '-\phi '\right)\right)+4 a^4 \mathcal{H}^2 f' \left(-2 a^2 C_0 k^2 \phi +4 a^4 k^2 \psi '-9 C_0^2 \psi '\right)+a^4 C_0 k^2 \mathcal{H} f' \left(C_0 \phi -8 a^2 \psi '\right)-72 a^2 C_0 \mathcal{H}^4 \left(2 a^4 \phi  f'+3 C_0 f'' \left(2 \phi '-\psi '\right)\right)-216 a^2 C_0 \mathcal{H}^5 f'' \left(8 a^2 \psi '+C_0 \phi \right)-1728 a^4 C_0 \mathcal{H}^6 \phi  f''\right)\Big)\,.
\end{dmath}
Finally, the combination
\begin{equation}
    \E_\phi\frac{C_0^2k^2-20C_0k^2\HH a^2+\HH^2(-36C_0^2+64k^2a^4)}{C_0^2k^2-8C_0k^2\HH a^2-144C_0\HH^3a^2-4\HH^2(9C_0^2-4k^2a^4)}+\E_j=0
\end{equation}
can be solved algebraically for $\psi'$,
\begin{dmath}
    {\psi'=\ }\frac{1}{C_0-4 a^2 \mathcal{H}}\Big(4 a^2 \mathcal{H}^2 \phi +C_0 \phi '\Big)\,,
\end{dmath}
upon which $\E_\phi=0$, and all perturbation equations have been solved. For this specific choice of $C$, we find the absence of scalar degrees of freedom, leading to vanishing Bardeen potentials $\Phi=\Psi=0$, implying the complete vanishing of scalar perturbations.
In the vector sector, the equations and solutions mirror those of connection 1, resulting in $V^i=0$, indicating the absence of vector perturbations.
Moving to the tensor sector, the equations of motion for $h_{ij}$ are like those of connection 1 with vacuum. Here, we observe two degrees of freedom for the tensor perturbations.

In total we thus have for general $C$ three scalar, two vector, and two tensor degrees of freedom, seven in total. Note that with our choice of solving for perturbations it is precisely the spatial metric (with six components) and the lapse (one component) that are dynamical. The shift variables $j,S^i$ are determined algebraically by the other variables. Even though no strong coupling is present, the way how the scalar and vector perturbations enter the kinetic matrix makes that either the scalar or the vector modes become a ghost. We show that explicitly for the more general case of curved space-times in section \ref{curvedfQcosmology} and taking the flat limit does not alter this fact. For the choice $C=C_0/a^2$, leading to de Sitter dynamics, we have no scalar and vector, and only two tensor degrees of freedom, five less than for general $C$. This indicates strong coupling in the de Sitter background in $f(\Q)$ cosmology for this background connection in both the scalar and vector sectors.

\subsubsection{Connection 3}\label{sssec:fQconn3}
We also first keep $C$ arbitrary leading to backgrounds that are in general not de Sitter space. Solving $\E_j=0$ for $j$,
\begin{dmath}[spread={5pt}]
    {j=\ }\frac{1}{3 k^2 \left(C'+2 C^2-2 C \mathcal{H}\right)^2}\\\Big(C \left(-\frac{a^2 f' \left(-9 \psi  C'-3 \phi  C'+4 C k^2 \sigma '-12 C \psi '-18 C \mathcal{H} (\psi +\phi )+12 \mathcal{H}^2 (3 \psi +\phi )\right)}{f''}+\frac{a^4 f (3 \psi +\phi )}{f''}+36 \mathcal{H}^2 \phi  \left(-C'-2 C^2+2 C \mathcal{H}\right)+36 \mathcal{H} \psi ' \left(-C'-2 C^2+2 C \mathcal{H}\right)+9 \left(C'+2 C^2-2 C \mathcal{H}\right) \left(2 \phi  C'+C \left(3 \psi '+\phi '\right)\right)-36 C \mathcal{H} \phi  \left(-C'-2 C^2+2 C \mathcal{H}\right)\right)\Big)\,,
\end{dmath}
we find that the kinetic matrix of the remaining $\E_\phi,\E_\psi,\E_\sigma$ has only rank two. The combination
\begin{equation}\label{eq:fQConn3Equ1}
    \frac{k^2(2\HH C-C^2+C')}{3C^2}\E_\phi+\frac{k^2}{3}\E_\psi+\E_\sigma
\end{equation}
is a first order differential equation only, and may be solved for $\phi'$ algebraically, which we call solution I, $\phi_I'$. The expression for it and the following formulas are very lengthy and we skip them since they are not necessary for the argument. Replacing $\psi\to\psi+\frac{k^2}{3}\sigma$ we can form the combination
\begin{equation}
    \frac{3\HH a^2 f-3(\HH(12\HH^2-12\HH C+C^2)+2(C-3\HH)C')f'}{3\HH ((C-2\HH)^2-2C')f'-a^2\HH f}\E_\phi-\E_\psi\,,
\end{equation}
which is now algebraic in $\phi$ and can be solved for it; we call this solution II, $\phi_{II}$. Now $\E_\sigma=0$, so we are left with $\E_\psi$. However, we also need to ensure that our two solutions agree, so we also keep \eqref{eq:fQConn3Equ1} with $\phi=\phi_{II}$ plugged in. Together with $\E_\psi$ we then have two second order differential equations in $\psi,\sigma$. Their kinetic matrix is very complicated, so we only note that $\det(K_{ij})\propto f''$, so $f''\neq 0$ is needed for the scalars to be fully dynamical. We have two scalar degrees of freedom.


In the vector sector solving $\E_{S^i}=0$ algebraically for $S^i={F^i}'$ we find $\E_{F^i}=0$ and $V^i=0$, so we have no vector degrees of freedom for any choice of $C$, as for connection 1.
In the tensor sector the equations for $h_{ij}$ read as for connection 1
\begin{align}
    \frac{f'}{2a^2}\left(h_{ij}''+(2\HH+\ln(f'(\Q))')h_{ij}'+(k^2+2u^2)h_{ij}\right)=0\,,
\end{align}
so we still have two tensor degrees of freedom but they have a distinctive feature from GR. Note, that this connection choice suffers like connection 1 from strong coupling since three of the gravity modes loose their kinetic terms for generic non-de Sitter cosmologies. 

For the special case $C=C_0/a^2$ we have also de Sitter like background dynamics. In the scalar sector we start from the beginning by solving $\E_j=0$ for $j$
\begin{dmath}
    j=\frac{1}{6 k^2 f'' \left(C_0-2 a^2 \mathcal{H}\right){}^2}\Big(36 a^2 \mathcal{H}^2 f'' \left(2 a^2 \psi '-C_0 \phi \right)+6 \mathcal{H} \left(a^6 \phi  f'-3 a^2 C_0 f'' \left(5 \psi '+\phi '\right)\right)-2 a^6 f' \left(k^2 \sigma '-3 \psi '\right)+72 a^4 \mathcal{H}^3 \phi  f''+9 C_0^2 f'' \left(3 \psi '+\phi '\right)\Big)\,,
\end{dmath}
and then $\E_\sigma+\frac{k^2}{3}(\E_\psi-\E_\phi)=0$ algebraically for $\phi'$. We call this solution $\phi_I$, and it has the form
\begin{dmath}[spread={5pt}]
    \phi'_I=\frac{1}{9 C_0 \mathcal{H} f'' \left(C_0-2 a^2 \mathcal{H}\right)}\Big(-2 \mathcal{H}^2 \left(3 a^2 C_0 f'' \left(2 k^2 \sigma '-9 \psi '\right)+4 a^4 k^2 f'' \left(k^2 \sigma -3 \psi \right)+3 a^6 \phi  f'-9 C_0^2 \phi  f''\right)+\mathcal{H} \left(8 a^2 C_0 k^2 f'' \left(k^2 \sigma -3 \psi \right)+2 a^6 f' \left(k^2 \sigma '-3 \psi '\right)-9 C_0^2 f'' \psi '\right)+24 a^4 \mathcal{H}^3 f'' \left(k^2 \sigma '-3 \psi '\right)-72 a^4 \mathcal{H}^4 \phi  f''-2 C_0^2 k^2 f'' \left(k^2 \sigma -3 \psi \right)\Big)\,.
\end{dmath}
Shifting $\psi\to\psi+\frac{k^2}{3}\sigma$ one can then solve $\E_\psi-3\E_\phi=0$ algebraically for $\phi_{II}=-\psi/\HH$, leading to $\E_\sigma=-\frac{k^2}{3}\E_\psi$. We then solve $\E_\phi=0$ algebraically for $\sigma'$,
\begin{equation}
    \sigma'=\frac{1}{C_0 k^2 \mathcal{H}}\Big(2 k^2 \psi  \left(C_0-2 a^2 \mathcal{H}\right)+3 C_0 \left(\psi ''-4 \mathcal{H} \psi '\right)\Big)\,,
\end{equation}
such that now $\E_\sigma=\E_\psi=0$. In this case we have $\phi_{II}=\phi_{I}$, so the two solutions are consistent. The final form of the perturbations is
\begin{dmath}
    {j=\ }\frac{1}{6 k^2 f'' \left(C_0-2 a^2 \mathcal{H}\right){}^2}\Big(36 a^2 \mathcal{H}^2 f'' \left(2 a^2 \left(\frac{2 k^2 \psi  \left(C_0-2 a^2 \mathcal{H}\right)+3 C_0 \left(\psi ''-4 \mathcal{H} \psi '\right)}{3 C_0 \mathcal{H}}+\psi '\right)+\frac{C_0 \psi '}{\mathcal{H}}\right)-2 a^6 f' \left(\frac{2 k^2 \psi  \left(C_0-2 a^2 \mathcal{H}\right)+3 C_0 \left(\psi ''-4 \mathcal{H} \psi '\right)}{C_0 \mathcal{H}}-3 \left(\frac{2 k^2 \psi  \left(C_0-2 a^2 \mathcal{H}\right)+3 C_0 \left(\psi ''-4 \mathcal{H} \psi '\right)}{3 C_0 \mathcal{H}}+\psi '\right)\right)+9 C_0^2 f'' \left(3 \left(\frac{2 k^2 \psi  \left(C_0-2 a^2 \mathcal{H}\right)+3 C_0 \left(\psi ''-4 \mathcal{H} \psi '\right)}{3 C_0 \mathcal{H}}+\psi '\right)+\psi '-\frac{\psi ''}{\mathcal{H}}\right)+6 \mathcal{H} \left(-3 a^2 C_0 f'' \left(5 \left(\frac{2 k^2 \psi  \left(C_0-2 a^2 \mathcal{H}\right)+3 C_0 \left(\psi ''-4 \mathcal{H} \psi '\right)}{3 C_0 \mathcal{H}}+\psi '\right)+\psi '-\frac{\psi ''}{\mathcal{H}}\right)-\frac{a^6 f' \psi '}{\mathcal{H}}\right)-72 a^4 \mathcal{H}^2 f'' \psi '\Big)\,,\\
    {\phi=\ }-\frac{\psi}{\HH}\,,\\
    {\sigma'=\ }\frac{1}{C_0 k^2 \mathcal{H}}\Big(2 k^2 \psi  \left(C_0-2 a^2 \mathcal{H}\right)+3 C_0 \left(\psi ''-4 \mathcal{H} \psi '\right)\Big)\,,
\end{dmath}
after applying $\psi\to\psi+\frac{k^2}{3}\sigma$. All equations of motion are then solved by the above solutions for $j,\phi,\sigma'$. We then also have that the Bardeen potentials vanish, $\Phi=\Psi=0$, so all scalar perturbations have been integrated out.
The vector sector is as for connection 1 and 2, so no connection perturbations appear, $V^i=0$, and thus no degrees of freedom are present. The tensor perturbations are as in GR, with the equations of motion as for connection 1 in de Sitter space.

In total we have for general $C$ two scalar and two tensor degrees of freedom, four in total, while for $C=C_0/a^2$ only the two tensor ones remain. This also indicates strong coupling issues in the scalar sector in the de Sitter background for this connection.

\subsection{Curved $f(\Q)$ cosmology}\label{curvedfQcosmology}

In curved cosmological backgrounds symmetric teleparallism admits only a single connection, namely with $\Gamma^0{}_{00}=\frac{C'-u^2}{C}$, $\Gamma^0{}_{ij}=C\g_{ij}$, $\Gamma^i{}_{0j}=\Gamma^i{}_{j0}=-\frac{u^2}{C}\delta^i_j$, and $\Gamma^i{}_{jk}=\gamma^i{}_{jk}$ the only non-vanishing components. $C$ is again a free function. In the limit $u\to0$ this connection becomes what we called connection 2 in the flat case. The background equations of motion are non-trivial even in vacuum, so we set\footnote{Including the scalar field $\Xi$ and $\delta\Xi$ does not lead to qualitatively different results.} $\Xi=\delta\Xi=0$, and obtain
\begin{align}
    \Q&=\frac{3}{C^2a^2}\left(2\HH^2C^2+2\HH C(C^2-u^2)+u^2C'+C^2(C'+2u^2)\right)\,,\\
    \HH'&=\frac{1}{6C^2(C^2+u^2)}\Big[f(\Q)a^2C^2(C^2+2\HH C+3u^2)+3f'(\Q)\times\\
    &\times\left(2C(\HH C+u^2)(-3u^2\HH+(u^2-4\HH^2)C+\HH C^2+C^3)+(C^2+u^2)(3u^2+2\HH C+C^2)C'\right)\Big]\,,\\
    C''&=\frac{1}{9C(C^2+u^2)^2}\Big[9\Big(-2C(u^2+\HH C)(-u^4C+8\HH^3C^2+C^5-4\HH^2 C(C^2-3u^2)+\\
    &+\HH (u^2-3C^2)(3u^2+C^2))+(u^2+2\HH C-C^2)(C^2+u^2)(3u^2+2\HH C+C^2)C'+2u^2(C^2+u^2){C'}^2\Big)+\\
    &+\frac{3C^2a^2\Big((u^2+2\HH C-C^2)(-2\HH C {f'(\Q)}^2+f(\Q)f''(\Q)(3u^2+2\HH C+C^2))+{f'(\Q)}^2C'(C^2+u^2)\Big)}{f'(\Q)f''(\Q)}+\\
    &+\frac{f(\Q)C^4a^4}{f''(\Q)}\Big]\,.
\end{align}

We can also choose to go to spatially curved de Sitter space by having the function $C$ to fulfill
\begin{equation}\label{eq:fQCurveddS}
    C'=-2C\frac{2u^2C+\HH(C^2-u^2)}{C^2+u^2}\,,
\end{equation}
while keeping $C^2+u^2\neq0$; this is analogous to the $C=C_0/a^2$ choice we made to go to flat de Sitter, but in curved space we found no closed form solution for $C$ from \eqref{eq:fQCurveddS}. We then find the background equations
\begin{align}
    \Q&=6\frac{\HH^2+u^2}{a^2}\,,\\
    f(\Q)&=2\Q f'(\Q)\,,\\
    \HH'&=\HH^2+u^2\,.
\end{align}
From the last equation we can infer that $\Q$ is constant, as it was in flat de Sitter space.

Let us start with a general background. In the scalar sector we can solve $\E_j=0$ for $j$,
\begin{align}
    j&=\frac{1}{6 a^2 C^3 \mathit{f}' \left(\left(C^2+\mathit{u}^2\right) C'+2 \left(C^2-2 C \mathcal{H}+\mathit{u}^2\right) \left(C \mathcal{H}+\mathit{u}^2\right)\right)+2 a^4 C^5 \mathit{f}+3 C k^2 \left(C^2+\mathit{u}^2\right) \mathit{f}'' \left(C'+2 C \mathcal{H}+2 \mathit{u}^2\right)^2}\times\nonumber\\
    &\times\Big[-a^4 C^4 \mathit{f} (3 \psi +\phi )+a^2 C^2 \mathit{f}' \left(\left(C^2+\mathit{u}^2\right) \left(-3 C' (3 \psi +\phi )-4 C \left(k^2-3 \mathit{u}^2\right) \sigma '+12 C \psi '\right)+\right.\nonumber\\
    &\left.+6 C \mathcal{H} \left(C^2 (\phi -3 \psi )+2 C \mathcal{H} (3 \psi +\phi )+3 \mathit{u}^2 (\psi +\phi )\right)\right)+9 \left(C^2+\mathit{u}^2\right) \mathit{f}'' \left(C'+2 \left(C \mathcal{H}+\mathit{u}^2\right)\right)\times\nonumber\\
    &\times\left(C' \left(2 C^2 \psi -2 \mathit{u}^2 \phi \right)+C \left(\left(C^2+\mathit{u}^2\right) \phi '+\psi ' \left(-C^2+4 C \mathcal{H}+3 \mathit{u}^2\right)+4 \left(C \mathcal{H}+\mathit{u}^2\right) (C \psi +\mathcal{H} \phi )\right)\right)\Big]\,.
\end{align}
We then find that the kinetic matrix of the remaining variables $\phi,\psi,\sigma$ has rank three,
\begin{align}
    K_{ij}&=\frac{1}{b}\begin{pmatrix} c & d & e \\ d & -3h & l \\  e & l & m \end{pmatrix}\,,\\
    b&=6 a^2 C^2 \mathit{f}' \left(\left(C^2+\mathit{u}^2\right) C'+2 \left(C^2-2 C \mathcal{H}+\mathit{u}^2\right) \left(C \mathcal{H}+\mathit{u}^2\right)\right)+\nonumber\\
    &+2 a^4 C^4 \mathit{f}+3 k^2 \left(C^2+\mathit{u}^2\right) \mathit{f}'' \left(C'+2 C \mathcal{H}+2 \mathit{u}^2\right)^2\,,\\
    c&=-9 \left(C^2+\mathit{u}^2\right)^2 \mathit{f}'' \left(a^2 C^2 \mathit{f}+3 \mathit{f}' \left(\left(C^2+\mathit{u}^2\right) C'+2 \left(C^2-2 C \mathcal{H}+\mathit{u}^2\right) \left(C \mathcal{H}+\mathit{u}^2\right)\right)\right)\,,\\
    d&=9 \left(C^2+\mathit{u}^2\right) \mathit{f}'' \left(a^2 C^2 \mathit{f} \left(C^2-4 C \mathcal{H}-3 \mathit{u}^2\right)+\mathit{f}' \left(2 \left(C \mathcal{H}+\mathit{u}^2\right) \left(\left(C^2+\mathit{u}^2\right) \left(3 C^2+2 k^2-9 \mathit{u}^2\right)+\right.\right.\right.\nonumber\\
    &\left.\left.\left.+24 C^2 \mathcal{H}^2+6 C \mathcal{H} \left(\mathit{u}^2-3 C^2\right)\right)+\left(C^2+\mathit{u}^2\right) C' \left(3 C (C-4 \mathcal{H})+2 k^2-9 \mathit{u}^2\right)\right)\right)\,,\\
    e&=-6 k^2 \left(C^2+\mathit{u}^2\right)^2 \mathit{f}' \mathit{f}'' \left(k^2-3 \mathit{u}^2\right) \left(C'+2 C \mathcal{H}+2 \mathit{u}^2\right)\,,\\
    h&=4 a^4 C^4 \mathit{f} \mathit{f}'+a^2 C^2 \left(3 \mathit{f} \mathit{f}'' \left(C^2-4 C \mathcal{H}-3 \mathit{u}^2\right)^2+4 \left(\mathit{f}'\right)^2 \left(\left(C^2+\mathit{u}^2\right) \left(3 C'-2 k^2+6 \mathit{u}^2\right)-\right.\right.\nonumber\\
    &\left.\left.-12 C^2 \mathcal{H}^2+6 C \mathcal{H} \left(C^2-\mathit{u}^2\right)\right)\right)-3 \mathit{f}' \mathit{f}'' \left(-\left(C^2+\mathit{u}^2\right) \left(C'+2 \mathit{u}^2\right) \left(C^2 \left(3 C^2+4 k^2-18 \mathit{u}^2\right)+\right.\right.\nonumber\\
    &\left.\left.+2 k^2 C'-8 k^2 \mathit{u}^2+27 \mathit{u}^4\right)+12 C^2 \mathcal{H}^2 \left(C^2 \left(5 C^2+2 k^2-30 \mathit{u}^2\right)-4 \left(C^2+\mathit{u}^2\right) C'+2 k^2 \mathit{u}^2+13 \mathit{u}^4\right)+\right.\nonumber\\
    &\left.+2 C \mathcal{H} \left(4 \left(C^2+\mathit{u}^2\right) C' \left(3 C^2+k^2-9 \mathit{u}^2\right)+C^2 \left(C^2 \left(-3 C^2-4 k^2+45 \mathit{u}^2\right)+16 k^2 \mathit{u}^2-93 \mathit{u}^4\right)+\right.\right.\nonumber\\
    &\left.\left.+5 \left(4 k^2 \mathit{u}^4-9 \mathit{u}^6\right)\right)-192 C^3 \mathcal{H}^3 \left(C^2-2 \mathit{u}^2\right)+192 C^4 \mathcal{H}^4\right)\,,\\
    l&=-2 k^2 \left(C^2+\mathit{u}^2\right) \mathit{f}' \left(k^2-3 \mathit{u}^2\right) \left(4 a^2 C^2 \mathit{f}'+3 \mathit{f}'' \left(-C^2+4 C \mathcal{H}+3 \mathit{u}^2\right) \left(C'+2 C \mathcal{H}+2 \mathit{u}^2\right)\right)\,,\\
    m&=\frac{2}{3} k^2 \mathit{f}' \left(k^2-3 \mathit{u}^2\right) \left(2 a^4 C^4 \mathit{f}+2 a^2 C^2 \mathit{f}' \left(\left(C^2+\mathit{u}^2\right) \left(3 C'+2 k^2\right)-12 C^2 \mathcal{H}^2+6 C \mathcal{H} \left(C^2-\mathit{u}^2\right)\right)+\right.\nonumber\\
    &\left.+3 k^2 \left(C^2+\mathit{u}^2\right) \mathit{f}'' \left(C'+2 C \mathcal{H}+2 \mathit{u}^2\right)^2\right)\,.
\end{align}
We thus have, similar to flat $f(\Q)$ cosmology with connection 2, three propagating scalar degrees of freedom, $\phi,\psi,\sigma$. The above kinetic matrix reduces to the one from the flat connection 2 in the limit $u\to0$. The determinant is now
\begin{align}
    \det(K_{ij})&=36b^2 k^2 \left(C^2+\mathit{u}^2\right)^2 \left(\mathit{f}'\right)^2 \mathit{f}'' \left(k^2-3 \mathit{u}^2\right) \left(a^2 C^2 \mathit{f}+3 \mathit{f}' \left(\left(C^2+\mathit{u}^2\right) C'-2 C \mathcal{H} \left(-C^2+2 C \mathcal{H}+\mathit{u}^2\right)\right)\right)\,.
\end{align}
This determinant vanishes in the case $C=\pm iu$, which is only viable for closed universes $u^2<0$. In this case one can check that an additional scalar mode may be integrated out, leading again to a strong coupling problem for this choice of (constant) $C$; we do not report the resulting kinetic matrix for the two remaining propagating scalar modes here. In the spatially flat case with $u=0$ this choice was not allowed as $C\neq0$.

In the vector sector we can solve $\E_{S^i}=0$ for $S^i$,
\begin{equation}
    S^i=\frac{3(k^2-2u^2)(C^2+u^2)f'}{-2fa^2C^2+3f'\left(8\HH^2C^2+4\HH C(u^2-C^2)+(C^2+u^2)(k^2-2u^2-2C')\right)}{F^i}'\,.
\end{equation}
We then find the equation of motion for $F^i$ to be
\begin{align}
    \E_{F^i}=&\frac{(k^2-2u^2)f'\Big(fa^2C^2+3f'(-2\HH C(u^2+2\HH C-C^2)+(u^2+C^2)C'\Big)}{a^2\Big(2fa^2C^2-3f'(8\HH^2C^2+4\HH C(u^2-C^2)+(C^2+u^2)(k^2-2u^2-2C'))\Big)}{F^i}''-\\
    &-\frac{1}{3 C \left(C^2+\mathit{u}^2\right) \left(2 a^3 C^2 \mathit{f}+3 a \mathit{f}' \left(\left(C^2+\mathit{u}^2\right) \left(2 C'-k^2+2 \mathit{u}^2\right)-8 C^2 \mathcal{H}^2+4 C \mathcal{H} \left(C^2-\mathit{u}^2\right)\right)\right)^2}\times\\
    &\times\Bigg(\left(k^2-2 \mathit{u}^2\right) \left(a^2 C^2 \mathit{f}+3 \mathit{f}' \left(\left(C^2+\mathit{u}^2\right) C'-2 C \mathcal{H} \left(-C^2+2 C \mathcal{H}+\mathit{u}^2\right)\right)\right) \times\\
    &\times\left(2 C^2 \left(a^2 C \mathit{f}-12 \mathcal{H} \mathit{f}' \left(C \mathcal{H}+\mathit{u}^2\right)\right) \left(a^2 C \mathit{f}-6 \mathcal{H} \mathit{f}' \left(-C^2+2 C \mathcal{H}+\mathit{u}^2\right)\right)+\right.\\
    &\left.+3 \left(C^2+\mathit{u}^2\right) C' \mathit{f}' \left(4 a^2 C^2 \mathit{f}+3 \mathit{f}' \left(-16 C^2 \mathcal{H}^2+4 C \mathcal{H} \left(C^2-3 \mathit{u}^2\right)+(\mathit{u}-C) (C+\mathit{u}) \left(k^2-2 \mathit{u}^2\right)\right)\right)+\right.\\
    &\left.+18 \left(C^2+\mathit{u}^2\right)^2 \left(C'\right)^2 \left(\mathit{f}'\right)^2\right)\Bigg){F^i}'-\\
    &-\frac{\left(k^2-2 \mathit{u}^2\right) \left(a^2 C^2 \mathit{f}+3 \mathit{f}' \left(\left(C^2+\mathit{u}^2\right) C'-2 C \mathcal{H} \left(-C^2+2 C \mathcal{H}+\mathit{u}^2\right)\right)\right)}{3 a^2 \left(C^2+\mathit{u}^2\right)}F^i\,.
\end{align}
Note that for vectors $k^2\geq 2u^2$, so the prefactor in front of ${F^i}''$ is non-vanishing unless $C$ is fine-tuned.  We have a second order equation of motion for $F^i$, hence two propagating vector modes $F^i$.
In the tensor sector we have the equation of motion
\begin{equation}
    \frac{f'}{2a^2}\left(h_{ij}''+(2\HH+\ln(f'(\Q))')h_{ij}'+(k^2+2u^2+2C\ln(f'(\Q))'h_{ij}\right)=0
\end{equation}
for $h_{ij}$, which has schematically the same form as in the flat case with connection 2.

In agreement with \cite{Gomes:2023tur} we can compute the ratio between the determinant of the kinetic matrix in the scalar sector, $K_s$, and the prefactor of ${F^i}''$ in $\E_{F^i}$, which we call $K_v$, in the ultraviolet limit,
\begin{equation}
    \frac{\det(K_s)}{K_v}\xrightarrow[]{k^2\to\infty}-\frac{18k^2f'{}^2(C^2+u^2)^2}{(C'+2\HH C+2u^2)^2}\,,
\end{equation}
which is strictly negative for generic connection function $C$ and any spatial curvature $u$. This shows that regardless of the choice of $f$, as long as all three scalar and both vector modes propagate, at least one scalar or the two vector modes will be a ghost for large momenta $k^2$, rendering this type of solutions unhealthy. The above formula also holds in the spatially flat case for connection 2 with $u=0$, so the ghost instability also occurs in flat geometries for generic background connections\footnote{Our formulas for $\det(K_s)$ and $K_v$ agree with those found in \cite{Gomes:2023tur} with the identification of connection functions
\begin{equation*}
    C=-\sigma_0\lambda/\xi'\,,
\end{equation*}
up to irrelevant numerical prefactors. }.

By going to de Sitter space the kinetic matrix $K_{ij}$ in the scalar sector has no longer full rank. We may solve the combination
\begin{align}
    \frac{-k^2(u^4+4u^2C^2+4\HH C^3-C^4)}{3(C^2+u^2)^2}\E_\phi+\frac{k^2}{3}\E_\psi+\E_\sigma=0
\end{align}
for $\phi'$, followed by solving
\begin{align}
    &\frac{16\HH^2C^3(C^4+u^4)+4u^2C(2C^6+13u^2C^4+4u^4C^2+u^6)-\HH(C^8-30u^2C^6+36C^4u^4+6C^2u^6+3u^8)}{\HH(C^2+u^2)^2(u^4+4u^2C^2+4\HH C^3-C^4)}\times\nonumber\\
    &\times\E_\phi+\E_\psi=0
\end{align}
for $\phi$. After putting $\psi\to\psi+\sigma k^2/3$ we find $\E_\sigma\propto\E_\psi$, and both solutions for $\phi$ agreeing after using the last remaining equation of motion $\E_\psi=0$. $\E_\psi$ is then a first order differential eqution in $\sigma$ - as well as second order in $\psi$ - so we can in principal solve it for $\sigma$. The final solutions are
\begin{align}
    j&=\frac{\psi}{\HH}+\sigma\,,\\
    \phi&=\frac{u^2\psi-\HH\psi'}{\HH^2}\,,\\
    \sigma'&=\frac{1}{k^2 \mathcal{H}^3 \left(-C^4+4 C^3 \mathcal{H}+4 C^2 u^2+u^4\right)}\times\\
    &\times\Big[-8 C^2 u^2 \mathcal{H}^4 \left(k^2 \sigma +3 \psi \right)-4 u^2 \mathcal{H}^3 \left(C \left(k^2 \sigma  \left(u^2-C^2\right)-\psi  \left(k^2-6 u^2\right)\right)+3
   \left(u^2-C^2\right) \psi '\right)+6 u^4 \psi  \left(C^2+u^2\right)^2+\nonumber\\
   &+\mathcal{H}^2 \left(2 u^2 \psi  \left(u^2-C^2\right) \left(-3 C^2+k^2+3 u^2\right)+3 \left(C^2+u^2\right)^2 \psi ''+24 C u^4 \psi '\right)+6 u^2 \mathcal{H}
   \left(-\left(C^2+u^2\right)^2 \psi '-4 C u^4 \psi \right)\Big]\,.\nonumber
\end{align}
We then also find $\Phi=\Psi=0$, so we have integrated out all scalar degrees of freedom.
In the vector sector we have the solutions $S^i={F^i}'$ from $\E_{S^i}$ and then $\E_{F^i}=0$ and $V^i=0$, so all vector modes can be integrated out.
In the tensor sector we have the GR-like equation
\begin{equation}
    \frac{f'}{2a^2}(h_{ij}''+2\HH h_{ij}'+(k^2+2u^2)h_{ij})=0\,.
\end{equation}
We conclude that we encounter the same strong coupling problem in spatially curved de Sitter space that we already found in flat de Sitter for connection 2, namely the disappearance of the extra scalar and vector degrees of freedom we found in general backgrounds. In fact the same modes as for connection 2 are the ones that propagate when we are away from flat or curved de Sitter space, owing to the similarity of the background connections.\\

In conclusion, in the spatially flat case for connection 1 the background is exclusively characterized by de Sitter dynamics in vacuum with $\Xi=0$, and analogous to $f(G)$ and $f(\T)$ cosmology the scalar and vector perturbations vanish, leaving only the two tensor modes $h_{ij}$ to propagate, signaling strong coupling problem. However, for general backgrounds with $\Xi\neq0$ we indeed find two more propagating scalar modes, leading to four propagating gravitational modes in total (two scalars and two tensors). The scenario also becomes more intriguing for connections 2 and 3, where even in vacuum the connection function $C$ influences the background equations, resulting in a more general spacetime background that is not confined to de Sitter. Consequently, we observe additional propagating degrees of freedom, with connection 2 having seven (three scalars, two vectors, and two tensors) and connection 3 having four (two scalars and two tensors). Notably, de Sitter dynamics can still be achieved with connections 2 and 3 by selecting $C\propto a^{-2}$, in which case only the two tensor modes survive. This reduction in propagating degrees of freedom in the de Sitter background points to potential strong coupling issues in de Sitter space for $f(\Q)$ cosmology, akin to strong coupling problems observed in $f(\T)$ gravity \cite{Bahamonde:2022ohm,PhysRevD.103.024054}. Similar issues also appeared for the spatially curved geometry, where we only had a single connection akin to connection 2 from the flat situation, with similar results. In general backgrounds we found seven propagating degrees of freedom (three scalars, two vectors, and two tensors), which reduced to the two tensor modes when going to curved de Sitter space by an appropriate choice of the function $C$. The strong coupling issues of de Sitter space persist in curved de Sitter as well. In addition, we find the for the spatially flat case with connection 2, and also for the spatiall curved geometry, at least one ghost is present in either the scalar or vector sector as long as $C$ is kept general and all seven degrees of freedom propagate. The connection choices 1 and 3 suffer also from strong coupling for generic non-de Sitter cosmologies, where only four out of seven degrees of freedom propagate. 
\section{Discussion}


In this paper, we have conducted an in-depth examination of first-order perturbation theory within teleparallel gravity theories. Our investigation encompassed the derivation of the general form of connection perturbations for general teleparallel, metric teleparallel, and symmetric teleparallel theories, elucidating their gauge transformations. Additionally, we introduced the scalar-vector-tensor splittings of the connection perturbations, enabling their application in cosmological scenarios. Our findings revealed that general flat connections entail 16 perturbations (in four spacetime dimensions), while metric teleparallel connections exhibit six perturbations, and symmetric teleparallel connections display only four. Of these, tensor perturbations exclusively appear in the most general case, with metric and symmetric teleparallism hosting four and two new vector modes, respectively. Notably, most of the connection perturbations manifest as scalars. Furthermore, we have identified a perturbative coincidence gauge in symmetric teleparallism, where the connection perturbations vanish entirely, facilitating the elimination of higher-order derivatives in connection perturbations.

As part of our investigation, we explored $f(G)$, $f(\T)$, and $f(\Q)$ cosmologies as illustrative examples. We thoroughly examined the permissible choices of background connections in each case. Remarkably, for most instances, the backgrounds turned out to be de Sitter in the absence of matter, except for spatially flat connections 2 and 3 and the spatially curved connection in $f(\Q)$ gravity. Here, the connection exhibited the potential to introduce more intriguing dynamics through a free function $C$ that influenced the cosmological background. Otherwise, the inclusion of matter is necessary to deviate from the de Sitter background in $f(G)$ and $f(\T)$ theories, which lead us to add a scalar field to discuss more general cosmological backgrounds.

However, we found that in $f(G)$ and $f(\T)$ theories we were still able to integrate out all gravitational scalar and vector modes, leaving us only with the single dynamical scalar mode of the scalar field, and the two tensor modes of the metric. In particular all perturbations of the teleparallel connection are either integrated out or undetermined. This holds both in de Sitter space and in more general backgrounds driven by the scalar, and also in both spatially flat and curved cases. We thus observe persistent strong coupling issues in theories of $f(G)$ and $f(\T)$. Moreover, the backgrounds and perturbations in $f(\T)$ gravity had the same form as in $f(G)$ gravity, since even though $f(G)$ gravity allows for more general background connections through the function $\K$, this function never appeared in either background or perturbation equations of motion.

The situation of $f(\Q)$ theory deviates slightly from this result, in that while the situation of having only two gravitational modes propagating in both flat and curved de Sitter space - achieved by removing the scalar field $\Xi$ and making appropriate choices for the free connection function $C$ - is the same, in more general backgrounds more dynamical degrees of freedom appeared, the number depending on the chosen background connection. We found in total four gravitational degrees of freedom for flat connections 1 and 3, and seven for flat connection 2 and the single curved connection. The apparent loss of these additional degrees of freedom in both flat and curved de Sitter space indicates a strong coupling problem of symmetric teleparallism in those backgrounds. Moreover, for connection 2 in the spatially flat case, or the single connection in the spatially curved background, we find as in \cite{Gomes:2023tur} that at least one of the propagating scalar or vector degrees of freedom is a ghost, rendering $f(\Q)$ gravity unhealthy for general connection functions $C$. 

These results render teleparallel cosmologies pathological either through strong coupling or ghost degrees of freedom. One possible way out would be to consider a direct coupling between the connection and matter fields (hence non-vanishing hypermomentum) or non-minimal couplings, but this is out of the scope of the current study. 

\section*{Acknowledgments}
LH is supported by funding from the European Research Council (ERC) under the European Unions Horizon
2020 research and innovation programme grant agreement No 801781. LH further acknowledges support
from the Deutsche Forschungsgemeinschaft (DFG, German Research Foundation) under Germany’s Excel-
lence Strategy EXC 2181/1 - 390900948 (the Heidelberg STRUCTURES Excellence Cluster).
MH gratefully acknowledges the full financial support by the Estonian Research Council through the Personal Research Funding project PRG356. The authors acknowledge networking support by the COST Actions CA18108 and CA21136.

\appendix
\section{Tetrad perturbations}\label{app:A}
In this work we presented the perturbations of the connection directly as
\begin{equation}
    \Gamma^\alpha{}_{\mu\nu}\to\Gamma^\alpha{}_{\mu\nu}+\delta\Gamma^\alpha{}_{\mu\nu}\,.
\end{equation}
However, in the literature one often finds the perturbations to be implemented in the tetrad formalism, for example \cite{Hohmann:2020vcv}, so we want to briefly show here how the two formalisms are related.

The tetrad $e_\mu{}^a$, with greek spacetime coordinate indices and latin local basis coordinate indices, is defined by
\begin{equation}
    e^\alpha{}_ae_{\alpha b}=\eta_{ab}\,,
\end{equation}
with $\eta_{ab}$ the $d$ dimensional Minkowski metric. Any teleparallel connection can then be written as
\begin{equation}\label{A:TetradConnection}
    \Gamma^\alpha{}_{\mu\nu}=e^\alpha{}_a\partial_\mu e_\nu{}^a\,,
\end{equation}
for some suitable tetrad, as one has $R^\alpha{}_{\beta\mu\nu}=0$ automatically for any $e_\mu{}^a$. We could also work with the tetrad instead of the connection. Perturbing the tetrad
\begin{equation}
    e_\mu{}^a\to e_\mu{}^a+\delta e_\mu{}^a\,
\end{equation}
leads to first order to
\begin{equation}
    \delta\Gamma^\alpha{}_{\mu\nu}=\partial_\mu e^\alpha{}_a\delta e_\nu{}^a-\Gamma^\beta{}_{\mu\nu}e^\alpha{}_a\delta e_\beta{}^a\,.
\end{equation}
Comparing this with $\delta\Gamma^\alpha{}_{\mu\nu}=\nabla_\mu\delta\Lambda^\alpha{}_\nu$ one can show that we have the simple identification
\begin{equation}
    \delta\Lambda^\mu{}_\nu=e^\mu{}_a\delta e_\nu{}^a\,,\, \delta e_\mu{}^a= e_\mu{}^a\delta\Lambda^\mu{}_\nu\,.
\end{equation}
The gauge transformation of the tetrad is then correctly
\begin{equation}
    \delta e_\mu{}^a\to\delta e_\mu{}^a+\xi^\nu\partial_\nu e_\mu{}^a-\partial_\mu\xi^\nu e_\nu{}^a\,.
\end{equation}

In metric-teleparallism we also have
\begin{equation}
    Q_{\alpha\mu\nu}=\partial_\alpha(g_{\mu\nu}-e_{\mu a}e_{\nu}{}^a)=0\,,
\end{equation}
so we need the tetrad to also fulfill
\begin{equation}
    e_{\mu a}e_\nu{}^a=g_{\mu\nu}\,.
\end{equation}
Perturbing this equation leads correctly to
\begin{equation}
    \delta\Lambda_{(\mu\nu)}=e_{(\mu}{}^a\delta e_{\nu)a}=\frac12\delta g_{\mu\nu}\,.
\end{equation}

In symmetric teleparallism there is another interesting way of writing the connection \cite{BeltranJimenez:2021auj}, namely as
\begin{equation}
    \Gamma^\alpha{}_{\mu\nu}=\frac{\partial x^\alpha}{\partial\zeta^\lambda}\partial_\mu\partial_\nu\zeta^\lambda\,,
\end{equation}
where $\zeta^\lambda$ is a collection of $d$ scalar functions such that $\partial x^\alpha/\partial \zeta^\lambda$ exists. We can then introduce perturbations as
\begin{equation}
    \zeta^\lambda\to\zeta^\lambda+\delta\zeta^\lambda\,,
\end{equation}
leading to
\begin{equation}
    \delta\Gamma^\alpha{}_{\mu\nu}=\frac{\partial x^\alpha}{\partial\zeta^\lambda}\partial_\mu\partial_\nu\delta\zeta^\lambda-\Gamma^\beta{}_{\mu\nu}\frac{\partial x^\alpha}{\partial\zeta^\lambda}\partial_\beta\delta\zeta^\lambda\,.
\end{equation}
This can be brought into contact with our previous definition of perturbations by
\begin{equation}
    \delta\rchi^\mu=\frac{\partial x^\mu}{\partial\zeta^\lambda}\delta\zeta^\lambda\,,\,\delta\zeta^\lambda=\frac{\partial\zeta^\lambda}{\partial x^\mu}\delta\rchi^\mu\,.
\end{equation}
For the gauge transformation we have
\begin{equation}
    \delta\zeta^\lambda\to\delta\zeta^\lambda-\xi^\mu\partial_\mu\zeta^\lambda\,,
\end{equation}
which is correct since the $\zeta^\lambda$ are scalars. Note that in the coincident gauge, $\Gamma^\alpha{}_{\mu\nu}=0$ or equivalently $\zeta^\lambda=x^\lambda$, we have $\delta\rchi^\mu=\delta\zeta^\mu$.
We can also link this representation of the perturbations to those of the tetrad by comparing with \eqref{A:TetradConnection},
\begin{equation}
    e_\mu{}^a=\partial_\mu\zeta^a\,.
\end{equation}
Perturbations are thus simply given by
\begin{equation}
    \delta e_\mu{}^a=\partial_\mu \delta\zeta^a\,.
\end{equation}
Comparing directly the tetrad to our definition of perturbations we can combine the above to
\begin{equation}
    \delta e_\mu{}^a=e_\nu{}^a\nabla_\mu\delta\rchi^\nu\,.
\end{equation}
Using the above formulae one may translate our results to perturbation theory performed in the tetrad formalism.

\bibliographystyle{utcaps.bst}
\bibliography{Bibliography}

\providecommand{\href}[2]{#2}\begingroup\raggedright\begin{thebibliography}{10}

\bibitem{Will:2005va}
C.~M. Will, ``{The Confrontation between general relativity and experiment},''
  \href{http://dx.doi.org/10.12942/lrr-2006-3}{{\em Living Rev. Rel.}
  {\bfseries 9} (2006) 3}, \href{http://arxiv.org/abs/gr-qc/0510072}{{\ttfamily
  arXiv:gr-qc/0510072}}.

\bibitem{Mukhanov:2005sc}
V.~Mukhanov, \href{http://dx.doi.org/10.1017/CBO9780511790553}{{\em {Physical
  Foundations of Cosmology}}}.
\newblock Cambridge University Press, Oxford, 2005.

\bibitem{LIGOScientific:2016aoc}
{\bfseries LIGO Scientific, Virgo} Collaboration, B.~P. Abbott {\em et~al.},
  ``{Observation of Gravitational Waves from a Binary Black Hole Merger},''
  \href{http://dx.doi.org/10.1103/PhysRevLett.116.061102}{{\em Phys. Rev.
  Lett.} {\bfseries 116} no.~6, (2016) 061102},
  \href{http://arxiv.org/abs/1602.03837}{{\ttfamily arXiv:1602.03837 [gr-qc]}}.

\bibitem{RevModPhys.61.1}
S.~Weinberg, ``The cosmological constant problem,''
  \href{http://dx.doi.org/10.1103/RevModPhys.61.1}{{\em Rev. Mod. Phys.}
  {\bfseries 61} (Jan, 1989) 1--23}.
  \url{https://link.aps.org/doi/10.1103/RevModPhys.61.1}.

\bibitem{SupernovaCosmologyProject:1998vns}
{\bfseries Supernova Cosmology Project} Collaboration, S.~Perlmutter {\em
  et~al.}, ``{Measurements of $\Omega$ and $\Lambda$ from 42 high redshift
  supernovae},'' \href{http://dx.doi.org/10.1086/307221}{{\em Astrophys. J.}
  {\bfseries 517} (1999) 565--586},
  \href{http://arxiv.org/abs/astro-ph/9812133}{{\ttfamily
  arXiv:astro-ph/9812133}}.

\bibitem{SupernovaSearchTeam:1998fmf}
{\bfseries Supernova Search Team} Collaboration, A.~G. Riess {\em et~al.},
  ``{Observational evidence from supernovae for an accelerating universe and a
  cosmological constant},'' \href{http://dx.doi.org/10.1086/300499}{{\em
  Astron. J.} {\bfseries 116} (1998) 1009--1038},
  \href{http://arxiv.org/abs/astro-ph/9805201}{{\ttfamily
  arXiv:astro-ph/9805201}}.

\bibitem{Clifton:2011jh}
T.~Clifton, P.~G. Ferreira, A.~Padilla, and C.~Skordis, ``{Modified Gravity and
  Cosmology},'' \href{http://dx.doi.org/10.1016/j.physrep.2012.01.001}{{\em
  Phys. Rept.} {\bfseries 513} (2012) 1--189},
  \href{http://arxiv.org/abs/1106.2476}{{\ttfamily arXiv:1106.2476
  [astro-ph.CO]}}.

\bibitem{Copeland:2006wr}
E.~J. Copeland, M.~Sami, and S.~Tsujikawa, ``{Dynamics of dark energy},''
  \href{http://dx.doi.org/10.1142/S021827180600942X}{{\em Int. J. Mod. Phys. D}
  {\bfseries 15} (2006) 1753--1936},
  \href{http://arxiv.org/abs/hep-th/0603057}{{\ttfamily arXiv:hep-th/0603057}}.

\bibitem{Heisenberg:2018vsk}
L.~Heisenberg, ``{A systematic approach to generalisations of General
  Relativity and their cosmological implications},''
  \href{http://dx.doi.org/10.1016/j.physrep.2018.11.006}{{\em Phys. Rept.}
  {\bfseries 796} (2019) 1--113},
  \href{http://arxiv.org/abs/1807.01725}{{\ttfamily arXiv:1807.01725 [gr-qc]}}.

\bibitem{Heisenberg:2014rta}
L.~Heisenberg, ``{Generalization of the Proca Action},''
  \href{http://dx.doi.org/10.1088/1475-7516/2014/05/015}{{\em JCAP} {\bfseries
  05} (2014) 015}, \href{http://arxiv.org/abs/1402.7026}{{\ttfamily
  arXiv:1402.7026 [hep-th]}}.

\bibitem{Heisenberg:2018acv}
L.~Heisenberg, ``{Scalar-Vector-Tensor Gravity Theories},''
  \href{http://dx.doi.org/10.1088/1475-7516/2018/10/054}{{\em JCAP} {\bfseries
  10} (2018) 054}, \href{http://arxiv.org/abs/1801.01523}{{\ttfamily
  arXiv:1801.01523 [gr-qc]}}.

\bibitem{Deffayet:2009wt}
C.~Deffayet, G.~Esposito-Farese, and A.~Vikman, ``{Covariant Galileon},''
  \href{http://dx.doi.org/10.1103/PhysRevD.79.084003}{{\em Phys. Rev. D}
  {\bfseries 79} (2009) 084003},
  \href{http://arxiv.org/abs/0901.1314}{{\ttfamily arXiv:0901.1314 [hep-th]}}.

\bibitem{Horndeski:1974ux}
G.~W. Horndeski, ``Second-order scalar-tensor field equations in a
  four-dimensional space,'' \href{http://dx.doi.org/10.1007/BF01807638}{{\em
  International Journal of Theoretical Physics} {\bfseries 10} no.~6, (1974)
  363--384}. \url{https://doi.org/10.1007/BF01807638}.

\bibitem{BeltranJimenez:2019}
J.~B. Jim\'{e}nez, L.~Heisenberg, and T.~S. Koivisto, ``{The Geometrical
  Trinity of Gravity},'' \href{http://dx.doi.org/10.3390/universe5070173}{{\em
  Universe} {\bfseries 5} no.~7, (2019) 173},
  \href{http://arxiv.org/abs/1903.06830}{{\ttfamily arXiv:1903.06830
  [hep-th]}}.

\bibitem{Hohmann:2022mlc}
M.~Hohmann, ``{Teleparallel Gravity},''
  \href{http://dx.doi.org/10.1007/978-3-031-31520-6_4}{{\em Lect. Notes Phys.}
  {\bfseries 1017} (2023) 145--198},
  \href{http://arxiv.org/abs/2207.06438}{{\ttfamily arXiv:2207.06438 [gr-qc]}}.

\bibitem{BeltranJimenez:2019odq}
J.~Beltr\'an~Jim\'enez, L.~Heisenberg, D.~Iosifidis, A.~Jim\'enez-Cano, and
  T.~S. Koivisto, ``{General teleparallel quadratic gravity},''
  \href{http://dx.doi.org/10.1016/j.physletb.2020.135422}{{\em Phys. Lett. B}
  {\bfseries 805} (2020) 135422},
  \href{http://arxiv.org/abs/1909.09045}{{\ttfamily arXiv:1909.09045 [gr-qc]}}.

\bibitem{BeltranJimenez:2017}
J.~Beltr\'{a}n~Jim\'{e}nez, L.~Heisenberg, and T.~Koivisto, ``{Coincident
  General Relativity},''
  \href{http://dx.doi.org/10.1103/PhysRevD.98.044048}{{\em Phys. Rev. D}
  {\bfseries 98} no.~4, (2018) 044048},
  \href{http://arxiv.org/abs/1710.03116}{{\ttfamily arXiv:1710.03116 [gr-qc]}}.

\bibitem{BeltranJimenez:2018vdo}
J.~Beltr\'an~Jim\'enez, L.~Heisenberg, and T.~S. Koivisto, ``{Teleparallel
  Palatini theories},''
  \href{http://dx.doi.org/10.1088/1475-7516/2018/08/039}{{\em JCAP} {\bfseries
  08} (2018) 039}, \href{http://arxiv.org/abs/1803.10185}{{\ttfamily
  arXiv:1803.10185 [gr-qc]}}.

\bibitem{Heisenberg:2023lru}
L.~Heisenberg, ``{Review on $f(Q)$ Gravity},''
  \href{http://arxiv.org/abs/2309.15958}{{\ttfamily arXiv:2309.15958 [gr-qc]}}.

\bibitem{BeltranJimenez:2019tme}
J.~Beltr\'an~Jim\'enez, L.~Heisenberg, T.~S. Koivisto, and S.~Pekar,
  ``{Cosmology in $f(Q)$ geometry},''
  \href{http://dx.doi.org/10.1103/PhysRevD.101.103507}{{\em Phys. Rev. D}
  {\bfseries 101} no.~10, (2020) 103507},
  \href{http://arxiv.org/abs/1906.10027}{{\ttfamily arXiv:1906.10027 [gr-qc]}}.

\bibitem{DAmbrosio:2020nev}
F.~D'Ambrosio, M.~Garg, and L.~Heisenberg, ``{Non-linear extension of
  non-metricity scalar for MOND},''
  \href{http://dx.doi.org/10.1016/j.physletb.2020.135970}{{\em Phys. Lett. B}
  {\bfseries 811} (2020) 135970},
  \href{http://arxiv.org/abs/2004.00888}{{\ttfamily arXiv:2004.00888 [gr-qc]}}.

\bibitem{Bajardi:2020wl}
F.~Bajardi, D.~Vernieri, and S.~Capozziello, ``Bouncing cosmology in f(Q)
  symmetric teleparallel gravity,''
  \href{http://dx.doi.org/10.1140/epjp/s13360-020-00918-3}{{\em The European
  Physical Journal Plus} {\bfseries 135} no.~11, (2020) 912}.
  \url{https://doi.org/10.1140/epjp/s13360-020-00918-3}.

\bibitem{Ayuso:2020dcu}
I.~Ayuso, R.~Lazkoz, and V.~Salzano, ``{Observational constraints on
  cosmological solutions of $f(Q)$ theories},''
  \href{http://dx.doi.org/10.1103/PhysRevD.103.063505}{{\em Phys. Rev. D}
  {\bfseries 103} no.~6, (2021) 063505},
  \href{http://arxiv.org/abs/2012.00046}{{\ttfamily arXiv:2012.00046
  [astro-ph.CO]}}.

\bibitem{Frusciante:2021sio}
N.~Frusciante, ``{Signatures of $f(Q)$-gravity in cosmology},''
  \href{http://dx.doi.org/10.1103/PhysRevD.103.044021}{{\em Phys. Rev. D}
  {\bfseries 103} no.~4, (2021) 044021},
  \href{http://arxiv.org/abs/2101.09242}{{\ttfamily arXiv:2101.09242
  [astro-ph.CO]}}.

\bibitem{Anagnostopoulos:2021ydo}
F.~K. Anagnostopoulos, S.~Basilakos, and E.~N. Saridakis, ``{First evidence
  that non-metricity f(Q) gravity could challenge \ensuremath{\Lambda}CDM},''
  \href{http://dx.doi.org/10.1016/j.physletb.2021.136634}{{\em Phys. Lett. B}
  {\bfseries 822} (2021) 136634},
  \href{http://arxiv.org/abs/2104.15123}{{\ttfamily arXiv:2104.15123 [gr-qc]}}.

\bibitem{Atayde:2021pgb}
L.~Atayde and N.~Frusciante, ``{Can $f(Q)$ gravity challenge $\Lambda$CDM?},''
  \href{http://dx.doi.org/10.1103/PhysRevD.104.064052}{{\em Phys. Rev. D}
  {\bfseries 104} no.~6, (2021) 064052},
  \href{http://arxiv.org/abs/2108.10832}{{\ttfamily arXiv:2108.10832
  [astro-ph.CO]}}.

\bibitem{DAmbrosio:2021pnd}
F.~D'Ambrosio, L.~Heisenberg, and S.~Kuhn, ``{Revisiting cosmologies in
  teleparallelism},'' \href{http://dx.doi.org/10.1088/1361-6382/ac3f99}{{\em
  Class. Quant. Grav.} {\bfseries 39} no.~2, (2022) 025013},
  \href{http://arxiv.org/abs/2109.04209}{{\ttfamily arXiv:2109.04209 [gr-qc]}}.

\bibitem{Hohmann:2021}
M.~Hohmann, ``{General covariant symmetric teleparallel cosmology},''
  \href{http://arxiv.org/abs/2109.01525}{{\ttfamily arXiv:2109.01525 [gr-qc]}}.

\bibitem{Capozziello:2022wgl}
S.~Capozziello and R.~D'Agostino, ``{Model-independent reconstruction of f(Q)
  non-metric gravity},''
  \href{http://dx.doi.org/10.1016/j.physletb.2022.137229}{{\em Phys. Lett. B}
  {\bfseries 832} (2022) 137229},
  \href{http://arxiv.org/abs/2204.01015}{{\ttfamily arXiv:2204.01015 [gr-qc]}}.

\bibitem{Dimakis:2022rkd}
N.~Dimakis, A.~Paliathanasis, M.~Roumeliotis, and T.~Christodoulakis, ``{FLRW
  solutions in f(Q) theory: The effect of using different connections},''
  \href{http://dx.doi.org/10.1103/PhysRevD.106.043509}{{\em Phys. Rev. D}
  {\bfseries 106} no.~4, (2022) 043509},
  \href{http://arxiv.org/abs/2205.04680}{{\ttfamily arXiv:2205.04680 [gr-qc]}}.

\bibitem{Esposito:2022omp}
F.~Esposito, S.~Carloni, and S.~Vignolo, ``{Bianchi type-I cosmological
  dynamics in gravity: a covariant approach},''
  \href{http://dx.doi.org/10.1088/1361-6382/ac9efd}{{\em Class. Quant. Grav.}
  {\bfseries 39} no.~23, (2022) 235014},
  \href{http://arxiv.org/abs/2207.14576}{{\ttfamily arXiv:2207.14576 [gr-qc]}}.

\bibitem{Zhao:2021zab}
D.~Zhao, ``{Covariant formulation of f(Q) theory},''
  \href{http://dx.doi.org/10.1140/epjc/s10052-022-10266-4}{{\em Eur. Phys. J.
  C} {\bfseries 82} no.~4, (2022) 303},
  \href{http://arxiv.org/abs/2104.02483}{{\ttfamily arXiv:2104.02483 [gr-qc]}}.

\bibitem{Lin:2021uqa}
R.-H. Lin and X.-H. Zhai, ``{Spherically symmetric configuration in $f(Q)$
  gravity},'' \href{http://dx.doi.org/10.1103/PhysRevD.103.124001}{{\em Phys.
  Rev. D} {\bfseries 103} no.~12, (2021) 124001},
  \href{http://arxiv.org/abs/2105.01484}{{\ttfamily arXiv:2105.01484 [gr-qc]}}.
  [Erratum: Phys.Rev.D 106, 069902 (2022)].

\bibitem{DAmbrosio:2021zpm}
F.~D'Ambrosio, S.~D.~B. Fell, L.~Heisenberg, and S.~Kuhn, ``{Black holes in
  f(Q) gravity},'' \href{http://dx.doi.org/10.1103/PhysRevD.105.024042}{{\em
  Phys. Rev. D} {\bfseries 105} no.~2, (2022) 024042},
  \href{http://arxiv.org/abs/2109.03174}{{\ttfamily arXiv:2109.03174 [gr-qc]}}.

\bibitem{Banerjee:2021mqk}
A.~Banerjee, A.~Pradhan, T.~Tangphati, and F.~Rahaman, ``{Wormhole geometries
  in $f(Q)$ gravity and the energy conditions},''
  \href{http://dx.doi.org/10.1140/epjc/s10052-021-09854-7}{{\em Eur. Phys. J.
  C} {\bfseries 81} no.~11, (2021) 1031},
  \href{http://arxiv.org/abs/2109.15105}{{\ttfamily arXiv:2109.15105 [gr-qc]}}.

\bibitem{Wang:2021zaz}
W.~Wang, H.~Chen, and T.~Katsuragawa, ``{Static and spherically symmetric
  solutions in f(Q) gravity},''
  \href{http://dx.doi.org/10.1103/PhysRevD.105.024060}{{\em Phys. Rev. D}
  {\bfseries 105} no.~2, (2022) 024060},
  \href{http://arxiv.org/abs/2110.13565}{{\ttfamily arXiv:2110.13565 [gr-qc]}}.

\bibitem{Parsaei:2022wnu}
F.~Parsaei, S.~Rastgoo, and P.~K. Sahoo, ``{Wormhole in f(Q) gravity},''
  \href{http://dx.doi.org/10.1140/epjp/s13360-022-03298-y}{{\em Eur. Phys. J.
  Plus} {\bfseries 137} no.~9, (2022) 1083},
  \href{http://arxiv.org/abs/2203.06374}{{\ttfamily arXiv:2203.06374 [gr-qc]}}.

\bibitem{Maurya:2022wwa}
S.~K. Maurya, K.~Newton~Singh, S.~V. Lohakare, and B.~Mishra, ``{Anisotropic
  Strange Star Model Beyond Standard Maximum Mass Limit by Gravitational
  Decoupling in f(Q)$f(Q)$ Gravity},''
  \href{http://dx.doi.org/10.1002/prop.202200061}{{\em Fortsch. Phys.}
  {\bfseries 70} no.~11, (2022) 2200061},
  \href{http://arxiv.org/abs/2208.04735}{{\ttfamily arXiv:2208.04735 [gr-qc]}}.

\bibitem{Hohmann:2020vcv}
M.~Hohmann, ``{General cosmological perturbations in teleparallel gravity},''
  \href{http://dx.doi.org/10.1140/epjp/s13360-020-00969-6}{{\em Eur. Phys. J.
  Plus} {\bfseries 136} no.~1, (2021) 65},
  \href{http://arxiv.org/abs/2011.02491}{{\ttfamily arXiv:2011.02491 [gr-qc]}}.

\bibitem{Jimenez:2018}
J.~Beltr\'{a}n~Jim\'{e}nez, L.~Heisenberg, and T.~S. Koivisto, ``{Teleparallel
  Palatini theories},''
  \href{http://dx.doi.org/10.1088/1475-7516/2018/08/039}{{\em JCAP} {\bfseries
  1808} no.~08, (2018) 039},
\href{http://arxiv.org/abs/1803.10185}{{\ttfamily arXiv:1803.10185 [gr-qc]}}.

\bibitem{Heisenberg:2022mbo}
L.~Heisenberg, M.~Hohmann, and S.~Kuhn, ``{Homogeneous and isotropic cosmology
  in general teleparallel gravity},''
  \href{http://arxiv.org/abs/2212.14324}{{\ttfamily arXiv:2212.14324 [gr-qc]}}.

\bibitem{Hohmann:2017jao}
M.~Hohmann, L.~Jarv, and U.~Ualikhanova, ``{Dynamical systems approach and
  generic properties of $f(T)$ cosmology},''
  \href{http://dx.doi.org/10.1103/PhysRevD.96.043508}{{\em Phys. Rev. D}
  {\bfseries 96} no.~4, (2017) 043508},
  \href{http://arxiv.org/abs/1706.02376}{{\ttfamily arXiv:1706.02376 [gr-qc]}}.

\bibitem{DAmbrosio:2021}
F.~D'Ambrosio, S.~D.~B. Fell, L.~Heisenberg, and S.~Kuhn, ``{Black holes in
  $f(\mathbb Q)$ Gravity},'' \href{http://arxiv.org/abs/2109.03174}{{\ttfamily
  arXiv:2109.03174 [gr-qc]}}.

\bibitem{DAmbrosio2021:2109.04209v2}
F.~D'Ambrosio, L.~Heisenberg, and S.~Kuhn, ``{Revisiting cosmologies in
  teleparallelism},'' \href{http://dx.doi.org/10.1088/1361-6382/ac3f99}{{\em
  Class. Quant. Grav.} {\bfseries 39} no.~2, (2022) 025013},
  \href{http://arxiv.org/abs/2109.04209}{{\ttfamily arXiv:2109.04209 [gr-qc]}}.

\bibitem{BeltranJimenez:2020sih}
J.~Beltr\'an~Jim\'enez, L.~Heisenberg, and T.~Koivisto, ``{The coupling of
  matter and spacetime geometry},''
  \href{http://dx.doi.org/10.1088/1361-6382/aba31b}{{\em Class. Quant. Grav.}
  {\bfseries 37} no.~19, (2020) 195013},
  \href{http://arxiv.org/abs/2004.04606}{{\ttfamily arXiv:2004.04606
  [hep-th]}}.

\bibitem{Bahamonde:2022ohm}
S.~Bahamonde, K.~F. Dialektopoulos, M.~Hohmann, J.~Levi~Said, C.~Pfeifer, and
  E.~N. Saridakis, ``{Perturbations in non-flat cosmology for f(T) gravity},''
  \href{http://dx.doi.org/10.1140/epjc/s10052-023-11322-3}{{\em Eur. Phys. J.
  C} {\bfseries 83} no.~3, (2023) 193},
  \href{http://arxiv.org/abs/2203.00619}{{\ttfamily arXiv:2203.00619 [gr-qc]}}.

\bibitem{1986ApJ}
L.~F. {Abbott} and R.~K. {Schaefer}, ``{A General, Gauge-invariant Analysis of
  the Cosmic Microwave Anisotropy},''
  \href{http://dx.doi.org/10.1086/164525}{{\em Astrophys. J.} {\bfseries 308}
  (Sept., 1986) 546}.

\bibitem{Gomes:2023tur}
D.~A. Gomes, J.~B. Jim\'enez, A.~J. Cano, and T.~S. Koivisto, ``{On the
  pathological character of modifications of Coincident General Relativity:
  Cosmological strong coupling and ghosts in $f(\mathbb{Q})$ theories},''
  \href{http://arxiv.org/abs/2311.04201}{{\ttfamily arXiv:2311.04201 [gr-qc]}}.

\bibitem{PhysRevD.103.024054}
J.~B. Jim\'enez, A.~Golovnev, T.~Koivisto, and H.~Veerm\"ae, ``Minkowski space
  in $f(T)$ gravity,''
  \href{http://dx.doi.org/10.1103/PhysRevD.103.024054}{{\em Phys. Rev. D}
  {\bfseries 103} (Jan, 2021) 024054}.
  \url{https://link.aps.org/doi/10.1103/PhysRevD.103.024054}.

\bibitem{BeltranJimenez:2021auj}
J.~Beltr\'an~Jim\'enez and T.~S. Koivisto, ``{Accidental gauge symmetries of
  Minkowski spacetime in Teleparallel theories},''
  \href{http://dx.doi.org/10.3390/universe7050143}{{\em Universe} {\bfseries 7}
  no.~5, (2021) 143}, \href{http://arxiv.org/abs/2104.05566}{{\ttfamily
  arXiv:2104.05566 [gr-qc]}}.

\end{thebibliography}\endgroup

\end{document}